\documentclass[]{aa}
\usepackage{siunitx}
\usepackage{graphicx}   
\usepackage[hyperfootnotes=true, hidelinks, colorlinks, citecolor=blue, linkcolor=blue, linktocpage, bookmarks=true]{hyperref}
\usepackage{txfonts}
\usepackage{soul}
\usepackage[dvipsnames]{xcolor}
\usepackage{multicol, blindtext}
\usepackage{multirow}

\newcommand{\kms}{km\,s$^{-1}$}

\newcommand{\hi}{\mbox{H\,{\sc i}}}

\newcommand{\oiii}{\mbox{[O\,{\sc iii}]}}
\newcommand{\nii}{\mbox{[N\,{\sc ii}]}}
\newcommand{\sii}{\mbox{[S\,{\sc ii}]}}
\newcommand{\oi}{\mbox{[O\,{\sc i}]}}
\newcommand{\hei}{\mbox{He\,{\sc i}}}
\newcommand{\neiii}{\mbox{[Ne\,{\sc iii}]}}
\newcommand{\oii}{\mbox{[O\,{\sc ii}]}}
\newcommand{\nev}{\mbox{[Ne\,{\sc v}]}}
\newcommand{\nodata}{--}

\begin{document} 

   \title{\hi\, 21-cm absorption in low- and high-excitation radio-loud AGNs at $z<0.5$ from MALS}
   \titlerunning{\hi\, absorption in low and high excitation radio galaxies}
   \authorrunning{Deka, P. P. et al.}

   \author{P. P. Deka \href{https://orcid.org/0000-0001-9174-1186}{\includegraphics[width=8pt]{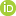}}
          \inst{1,2,3}
          \and
          N.~Gupta \href{https://orcid.org/0000-0001-7547-4241}{\includegraphics[width=8pt]{orcid.png}}\inst{1}
          \and
          J-.K.~Krogager\inst{4,5}
          \and
          S. A. Balashev\inst{6}
          \and
          H.-W. Chen\inst{7}
          \and
          F. Combes\inst{8}
          \and
          H.-R. Kl\"ockner\inst{9}
          \and
          P. Noterdaeme\inst{10}
          }

   \institute{Inter-University Centre for Astronomy and Astrophysics, Post Bag 4, Ganeshkhind, Pune 411 007, India
        \and
            Department of Astronomy, University of Cape Town, Private Bag X3,
Rondebosch 7701, South Africa\\
            \email{\href{partha.deka@uct.ac.za}{partha.deka@uct.ac.za}}
        \and
            SARAO, 2 Fir Street, Black River Park, Observatory 7925, South Africa
        \and
            Universit\'e Lyon1, ENS de Lyon, CNRS, Centre de Recherche Astrophysique de Lyon UMR5574, F-69230 Saint-Genis-Laval, France
        \and
            French-Chilean Laboratory for Astronomy, IRL 3386, CNRS and Universidad de Chile, Santiago, Chile
        \and
            Ioffe Institute, 26 Politeknicheskaya st., St. Petersburg, 194021, Russia
        \and
            Department of Astronomy and Astrophysics, The University of Chicago, 5640 S. Ellis Avenue, Chicago, IL 60637, USA
        \and
            Observatoire de Paris, Coll\`ege de France, PSL University, Sorbonne University, CNRS, LUX, Paris, France
        \and
            Max-Planck-Institut f\"ur Radioastronomie, Auf dem H\"ugel 69, D-53121 Bonn, Germany
        \and
            Institut d'Astrophysique de Paris, UMR 7095, CNRS-SU, 98bis bd Arago, 75014  Paris, France
            }
    \date{Received \today; accepted bbb}

\abstract
{
We present results from a search of cold neutral gas associated with radio-loud active galactic nuclei (AGNs) at $z<0.5$ -- corresponding to a lookback time of $\sim 36$\% of the age of the Universe -- using \hi\ 21-cm absorption line measurements from the MeerKAT Absorption Line Survey (MALS). 
Cross-matching the MALS 1006\,MHz catalog with the SDSS DR18 spectroscopic catalog yields 1908 radio sources at $z<0.5$. Of these, 613 are classified as AGNs using Baldwin-Phillips-Terlevich (BPT) diagnostics and radio luminosity criteria. We further classify 426 AGNs into 327 low-excitation radio galaxies (LERGs) and 99 high-excitation radio galaxies (HERGs) using a framework combining multiple optical emission-line diagnostics. We find a significant ($>3\sigma$) difference in $k$-corrected $g-r$ color, consistent with LERGs residing in older galaxies with quenched star formation. 
We search a radio-bright subsample of 79 LERGs and 20 HERGs ($S_{\rm 1.4\,GHz} > 4$\,mJy) for associated \hi\ 21-cm absorption using MALS L-band spectra. This sub-sample spans nearly six decades in radio luminosity ($\log\,\rm L_{\rm 1.4GHz}$~(W\,Hz$^{-1}$) $\sim 21.1-27.0$), reaching luminosities an order of magnitude fainter than previous targeted \hi\ surveys.
We report five new detections (4 LERGs, 1 HERG) at $0.29<z<0.47$, yielding an overall detection rate of $3^{+3}_{-2}\%$ at a $3\sigma$ integrated optical depth sensitivity threshold of 10.0\,\kms. This rate is consistent with sensitivity-matched low-$z$ ($<0.2$) samples, suggesting no significant redshift evolution out to $z\sim 0.5$ or dependence on radio luminosity.
We quantify absorption-line properties using velocity offset, asymmetry, and width. In three systems, absorption is entirely redshifted relative to the systemic velocity; in the remaining two, it is predominantly blueshifted, indicating complex gas kinematics. \hi\ profiles associated with LERGs show a wide range of asymmetries and velocity offsets exceeding 350\,\kms, indicative of disturbed cold-gas kinematics likely driven by lobe expansion or jet activity.}

\keywords{quasars: absorption lines ---  galaxies: ISM}

\maketitle
%
\section{Introduction} 
\label{sec:intro}  

The population of radio-loud Active Galactic Nuclei (AGNs) exhibits two distinct classes based on the nature of accretion and the accreting gas \citep[see][for a review]{Heckman14}: {\it (i)} Low-Excitation Radio Galaxies (LERGs),  where the central supermassive black hole (SMBH) is believed to be fueled by accretion of hot halo gas at rates $<1\%$ of the Eddington rate through a geometrically thick but optically thin accretion disc; and {\it (ii)} High-Excitation Radio Galaxies (HERGs),  where the SMBH is known to accrete cold gas at higher rates, $>1\%$ of the Eddington rate,  through a geometrically thin but optically thick accretion disc. Observationally, they are distinguished based on the strength of high-excitation emission lines (e.g., \oiii\,) in their optical spectra relative to low-excitation lines, such as \nii\,, \sii\,, and \oi\,, as the radiatively efficient central AGNs in HERGs lead to stronger emission lines of higher ionization compared to LERGs. The corresponding host galaxies also exhibit fundamental differences, with LERGs primarily hosted by galaxies having higher stellar and black hole masses and older stellar populations compared to HERGs. In the nearby universe ($z<0.3$), LERGs display both Fanaroff and Riley I and II \citep[FRI/II;][]{Fanaroff1974} morphologies, while HERGs are predominantly FRII. Additionally, the luminosity function of LERGs indicates little to no redshift evolution, or even negative evolution, whereas HERGs exhibit strong and positive evolution  in their comoving space density across all radio powers \citep{Buttiglione10, Best12, Best2014}.

HERGs and LERGs also display fundamentally distinct feedback mechanisms \citep[see][and the references therein]{Best12, Pracy16, Singha2021}.  In HERGs, radiatively efficient accretion onto the SMBH produces powerful radiative or quasar-mode feedback, often manifesting as high-velocity winds and broad ionized outflows driven by energy or radiation pressure \citep{Mullaney13, Harrison2014}. These can couple strongly to the host galaxy’s ISM, heating, ionizing, and/or relocating large amounts of gas, thereby regulating star formation on galactic scales. In contrast, LERGs accrete via radiatively inefficient flows, leading to feedback dominated by kinetic or radio-mode processes. This typically involves collimated relativistic jets that interact with the surrounding medium, inflating cavities in the hot halo gas, suppressing cooling flows in clusters and groups, and driving multiphase gas outflows \citep{Best12, Gitti2012, Chandola2020, Morganti2017}. Such jet–ISM interactions can introduce strong kinematic disturbances into the cold gas, leaving it unsettled and turbulent, thereby inhibiting its ability to collapse and form stars \citep[see][]{Ishibashi2014, Morganti2017, Harrison2018}.
The  differing nature of feedback processes governing these two populations may be responsible for their different evolutionary tracks. However, observational evidence is mixed, and it remains challenging to distinctly identify the relative importance of different feedback modes operating on various scales.

The classification of AGN into HERGs and LERGs requires the  measurement of flux densities for various optical emission lines.  The most sophisticated  criterion is based on the excitation index (EI),
\begin{equation}
\begin{aligned}
    {\rm EI} = \log({\rm \oiii\,/H\beta}) ~ - ~\frac{1}{3}[\log({\rm \nii\,/H\alpha}) ~ + \\ \log({\rm \sii\,/H\alpha}) ~ + ~ \log({\rm \oi\,/H\alpha})]),
    \label{eq:exindex}
\end{aligned}
\end{equation}
utilizing four emission-line ratios \citep[][]{Buttiglione10}.  As demonstrated by \citet[][]{Buttiglione10} for a sample of  113 local ($z<0.3$) radio galaxies from the Third Cambridge Catalog of Radio Sources (3CR),  an EI threshold of 0.95 divides radio galaxies into distinct populations of HERGs and LERGs.
Naturally, obtaining reliable detections of all six emission lines needed for calculating the EI is often challenging. Therefore, a variety of classification mechanisms based on fewer lines have been devised, resulting in several large HERG/LERG catalogs in the literature.  Among the most notable of these,  with further details presented later in the paper, are the classifications of 18,286 and 6,720 radio galaxies by \citet[][]{Best12} and \citet[][]{Pracy16}, respectively.  

Observations of \hi\ 21-cm absorption lines can reveal the kinematics and distribution of cold atomic gas associated with HERGs and LERGs, providing insight into the distinguishing features of these two populations.
\citet[][]{Chandola2020} found that \hi\ 21-cm absorption detection rates among HERGs and LERGs depend crucially on the extent of the radio emission and the mid-infrared (MIR) colors measured by the Wide-field Infrared Survey Explorer \citep[WISE;][]{Wright10, Cutri14}. Using a sample of 191 LERGs and 27 HERGs at $z<0.25$, they  found comparable \hi\ 21-cm absorption detection rates  between the two types when matched in radio size and WISE color  ($60.6^{+16.8}_{-13.4}$\% for LERGs and $50.0^{+26.9}_{-18.4}$\% for HERGs).  This suggests  that at least a fraction of LERGs contain significant amounts of gas and dust.  It is likely that in  LERGs, this gas is more susceptible to feedback effects, hindering its transport to the vicinity of the central black hole.
However, the evidence remains sparse due to considerable statistical uncertainties, and the existing samples have been limited to lower redshifts \citep[$z<0.25$;][]{Maccagni17}.

\begin{figure*}[t]
  \centering
  \begin{minipage}[b]{0.72\textwidth}
    \includegraphics[width=\textwidth]{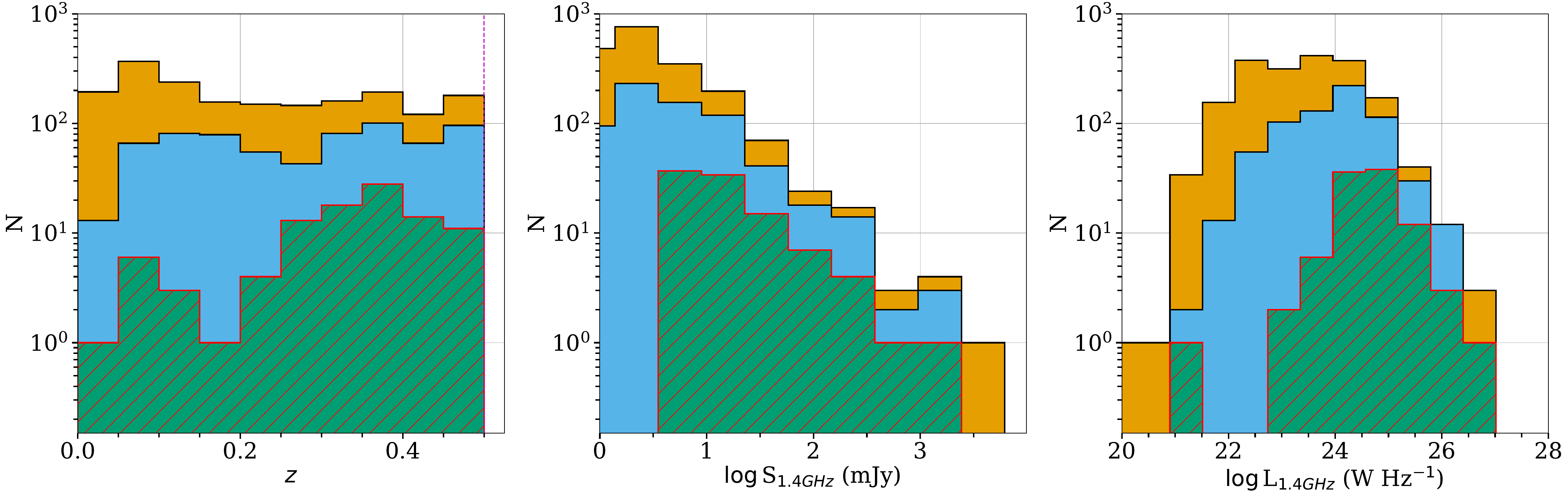}
  \end{minipage}
  \hfill
  \begin{minipage}[b]{0.245\textwidth}
    \includegraphics[width=\textwidth]{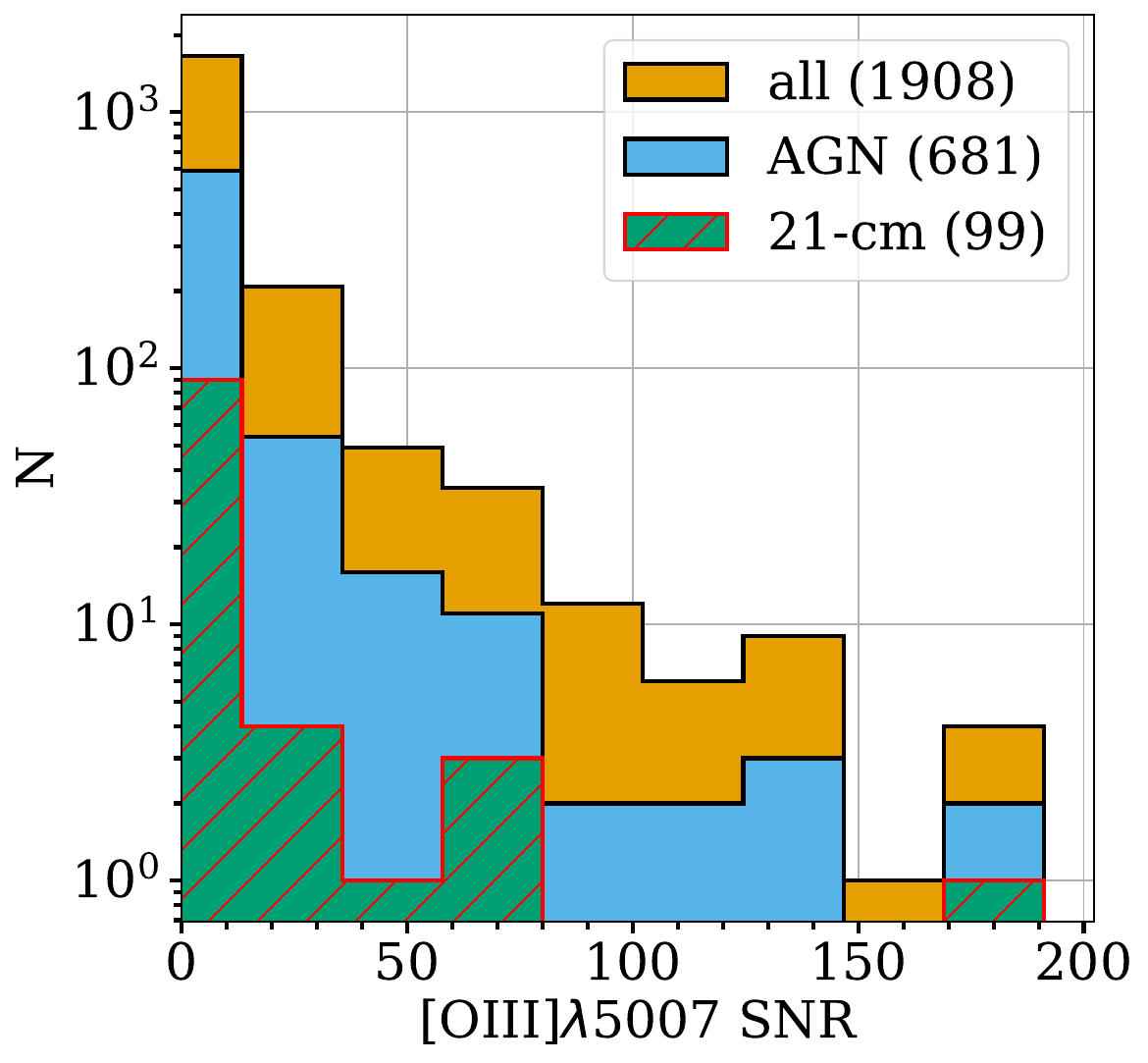}
  \end{minipage}
  \caption{Distribution of $z$ ({\it first panel}), S$_{1.4\,\rm GHz}$ ({\it second}),  L$_{1.4\rm GHz}$ ({\it third}) and S/N of \oiii$\lambda$5007 (\textit{fourth}) for the initial sample of 1908 radio sources (yellow histograms).  Of these, 681 are classified as AGNs (blue; see Sections~\ref{sec:GELATO} and \ref{sec:SF_AGN} for details).  Finally, 99 LERGs/HERGs from this subsample were searched for associated \hi\, 21-cm absorption (green hatched; details in Section~\ref{sec:21cm}). The low-$z$ deficit in the green histogram reflects the paucity of sources exceeding the 4\,mJy flux density cut applied for the 21-cm absorption search.
  }
  \label{fig:fig1}
\end{figure*}

In this paper, we use radio continuum images and spectra from the MeerKAT Absorption Line Survey \citep[MALS;][]{Gupta17mals} to extend \hi\ 21-cm line studies of radio galaxies to higher redshifts and lower radio luminosities.  While previous surveys of \hi\ absorption in LERGs and HERGs were limited to $z \lesssim 0.2$ \citep[e.g.,][]{Chandola2020, Maccagni17}, the present work extends this redshift range to $z \sim 0.5$, thereby providing the highest-$z$ sample of such galaxies studied to date.
Section~\ref{sec:sample} presents the MALS data products and cross-matching with the Sloan Digital Sky Survey (SDSS) data release 18 (DR18)  used to classify  1908 radio sources into HERGs and LERGs,  employing a framework that combines multiple optical-emission line diagnostics. Section~\ref{sec:21cm} presents the results of our \hi\ 21-cm search in the HERG/LERG sample, along with descriptions of new detections and their statistical properties.  We also discuss the kinematics of the absorbing gas and its implications for inflow and AGN feedback.
The results are summarized in Section~\ref{sec:summary_LERGHERG}.
Throughout  this paper, we adopt a $\Lambda$CDM cosmology with $\Omega_m$=0.315, $\Omega_\Lambda$=0.685, and H$_{\rm 0}$=67.4\,\kms\,Mpc$^{-1}$ \citep[][]{Planck20}. Unless otherwise stated, all logarithms ($\log$) are base 10.

\section{Sample: HERGs and LERGs} 
\label{sec:sample}

 This section details the cross-matching of MALS data release 1 (DR1) with SDSS DR18, and the HERG/LERG classification procedure using emission lines detected in the SDSS spectra.

\subsection{MALS observations and data products} 
\label{sec:obs}

MALS observed 391 pointings at L-band (900 -- 1670\,MHz) during the first phase of the survey from 2020, April, 01 to 2021, January, 18.  \citet[][]{Gupta22salt} presents the details of a large spectroscopic campaign to define the survey footprint \citep[see also][]{Krogager18}.  The MeerKAT data were processed using the Automated Radio Telescope Imaging Pipeline \citep[ARTIP;][]{Gupta21}, which is based on the NRAO Common Astronomy Software Applications (CASA) package \citep[][]{Casa22} to produce  continuum images and spectral line cubes for 15 spectral windows (SPWs) covering the L-band.
The details of continuum images and the associated catalog of 715,760 unique radio sources detected at a signal-to-noise ratio (S/N)\,$>$5 are presented in the MALS DR1 paper \citep[][]{Deka2024}.
Briefly, the median rms noise away from the pointing center is $\sim$20\,$\mu$Jy\,beam$^{‑1}$, and the spatial resolution across the band is $8^{\prime\prime}$ $-$ $12^{\prime\prime}$. 
The DR1 catalog includes spectral indices, $\alpha$, where $S_\nu \propto \nu^\alpha$, based on the flux densities measured at 1006\,MHz and 1381\,MHz. 
The L-band spectra in the heliocentric frame used here to search for \hi\ 21-cm absorption were extracted towards pixels corresponding to the peak flux density of the radio source.  The absorption was searched within $\pm 2000$~\kms\ of the redshifted \hi\ 21-cm line frequency following the methodology adopted for the MALS Galactic spectral line search \citep[][]{Gupta25galhi}.
Further details on the radio sources and their associated spectra are provided in subsequent sections.

\subsection{Cross-matching with SDSS}

We cross-matched the MALS DR1 catalog at 1006\,MHz with the SDSS DR18 spectroscopic catalog using a  search radius of 5$^{\prime\prime}$ to identify 1965 radio sources at $z<0.5$, the redshift range suitable to search for \hi\ 21-cm absorption using MeerKAT's L-band.   
A matching radius of 2$^{\prime\prime}$ should be optimal for cross-matching with SDSS \citep[][]{Lu07, Gupta21salt}. However, a 5$^{\prime\prime}$ radius was used to compensate for the large synthesized beam (12$^{\prime\prime}$) at 1006\,MHz. This ensures that AGNs with extended radio morphologies, such as FRI and FRII  sources, are not excluded. For 152 sources, the flux-weighted mean radio positions differed by more than $2^{\prime\prime}$, even though the peak radio positions remained consistent within that threshold. To ensure a reliable cross-matched sample, we visually inspected all matches obtained from the Panoramic Survey Telescope and Rapid Response System \citep[PanSTARRS;][]{Chambers16}. Notably, the 152 sources with separations between $2^{\prime\prime}$ and $5^{\prime\prime}$ predominantly exhibited extended morphologies. However, in all but 57 cases, the optical source coincided with the position of the radio core i.e., the radio component likely coincident with the AGN. After excluding these 57 spurious matches, our sample comprises 1908 sources.
The first three panels of Fig.~\ref{fig:fig1} show the distributions of redshift ($z$), 1.4\,GHz flux density (S$_{1.4\,\rm GHz}$) and luminosity at 1.4~GHz (L$_{1.4\,\rm GHz}$) for the sample. We note that 53/1908 sources were not detected at 1380~MHz.  For these 1006\,MHz flux densities were scaled to 1.4\,GHz using a typical spectral index of $\alpha = -0.7$.

\subsection{Spectroscopic fitting using GELATO} 
\label{sec:GELATO}

The LERGs/HERGs classification is carried out in two steps.  In Section~\ref{sec:SF_AGN}, we first use the Baldwin-Phillips-Terlevich (BPT) emission line diagnostic method \citep[][]{Baldwin1981} based on the location of the source in the \oiii/H$\beta$ $-$ \nii/H$\alpha$ plane,  to distinguish radio sources as either AGN or star-forming (SF) galaxies.  We also use H$\alpha$- line luminosity (L$_{\rm H\alpha}$) and L$_{1.4\,\rm GHz}$ for this purpose.  Both these luminosities serve as independent measures of the star-formation rate for SF-galaxies. Conversely, for radio-loud AGNs, an excess of radio emission over  that estimated from the H$\alpha$ luminosity is expected.
In the second step (Section~\ref{sec:LERG_HERG}), AGNs are classified into HERGs and LERGs.  Notably, across $0\leq z\leq 0.5$, the \oiii$\lambda 5007$ line remains within the wavelength coverage of the SDSS spectra. Consequently, the classification of objects into LERGs/HERGs is feasible using one of the several criteria defined by \citet[][]{Best12}, even though at the highest redshifts, the majority of  other lines, including \nii$\lambda 6583$, \sii$\lambda\lambda 6716, 6731$, \oi$\lambda 6300$, and H$\alpha$, are beyond the SDSS coverage.

\begin{centering}
\begin{table}
\centering
\caption{The GELATO Emission Line Dictionary used in this work. Emission lines are grouped into \textit{Species}, and \textit{Species} into \textit{Groups}. Lines belonging to a given \textit{Species} (e.g., \sii$\lambda 6716.44$ and \sii$\lambda 6730.82$]) must share common kinematics, namely, redshift and velocity dispersion. Additional components, {\tt Outflow} for \oiii\, and {\tt Broad} for Balmer lines, and the constraints imposed on their kinematical properties are discussed in Section~\ref{sec:GELATO}.} \label{tab:GELATO_PARAMS}
\begin{tabular}{ccc}
\hline\hline
\textbf{Species} & \textbf{Line $\lambda_{\rm air}$ [\AA]} & \textbf{Ratio} \\
\hline\hline
 \multicolumn{3}{c}{}\\
 \multicolumn{3}{c}{\textit{Narrow Group}}\\
  \multicolumn{3}{c}{}\\
\hline
\multirow{2}{*}{\sii\,} & 6716.44 & - \\
                       & 6730.82 & - \\
\hline
\multirow{2}{*}{\nii\,} & 6583.45 & 1 \\
                       & 6548.05 & 0.34 \\
\hline
\multirow{2}{*}{\oi\,}  & 6300.30 & 1 \\
                       & 6363.78 & 0.333 \\
\hline
\hei\,                    & 5875.62 & - \\
\hline
\multirow{3}{*}{\oiii\,}& 5006.84 & 1 \\
                       & 4958.91 & 0.35 \\
                       & 4363.21 & - \\
\hline
\neiii\,                & 3868.76 & - \\
\hline
\multirow{2}{*}{\oii\,} & 3726.03 & - \\
                       & 3728.82 & - \\
\hline
\nev\,                  & 3425.88 & - \\
\hline\hline
 \multicolumn{3}{c}{}\\
 \multicolumn{3}{c}{\textit{Balmer Group}}\\
  \multicolumn{3}{c}{}\\
\hline
\hi\   & 6562.79 & - \\
\hline
\hi   & 4861.28 & - \\
\hline
\hi   & 4340.47 & - \\
\hline\hline
\end{tabular}
\end{table}
\end{centering}


\begin{figure*}[t]
\centerline{\vbox{
\centerline{\hbox{ 
\includegraphics[
trim = {0cm 0cm 0cm 0cm}, clip=true,
width=\textwidth,angle=0]{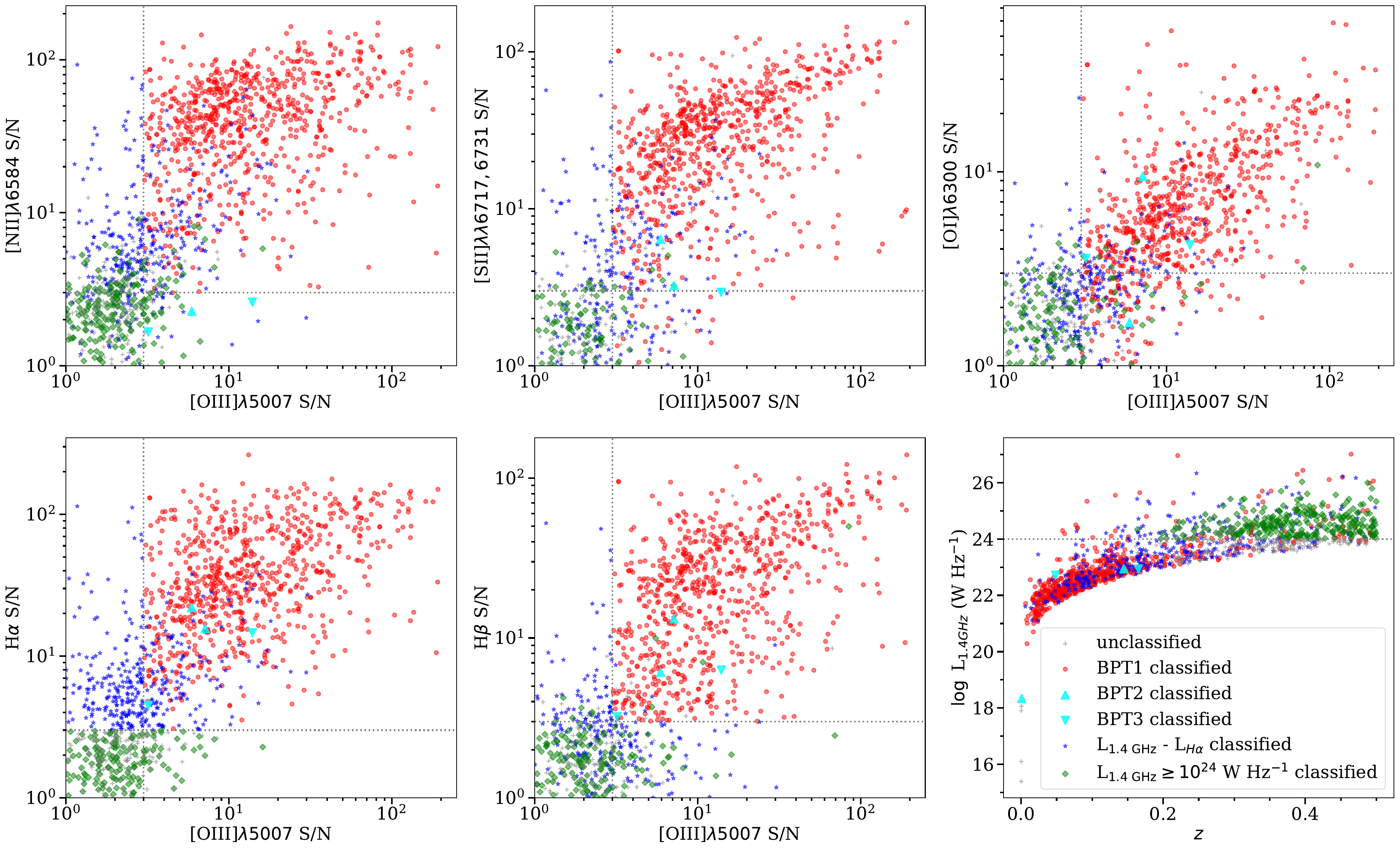}  
}} 
}}  
\vskip+0.0cm   
\caption{S/N distribution of various lines used for AGN / SF classification, and the location of sources in the L$_{1.4\,\rm GHz}$--$z$ plane based on the  classification method (details in Section~\ref{sec:SF_AGN}).  Sources with line S/N$<1$ are not shown. The dashed vertical and horizontal lines denote an S/N = 3, except in the bottom right panel where the dashed horizontal line (L$_{1.4\,\rm GHz}$ = $10^{24}$\,W\,Hz$^{-1}$) marks the boundary segregating SF and AGN-dominated sources according to \citet[][]{Pracy16}.
}
\label{fig:snr}
\end{figure*}

We used the Galaxy/AGN Emission-Line Analysis TOol\footnote{\href{https://github.com/TheSkyentist/GELATO}{https://github.com/TheSkyentist/GELATO}} \citep[GELATOv2.5.2;][]{Hviding2022} to measure the emission line fluxes, the equivalent widths and the associated uncertainties.  For a given spectrum with an initial redshift estimate ($z_0$) accurate to at least 0.5\%, GELATO fits a user-specified list of spectral lines to the input spectrum. 
The emission lines used in our analysis are summarized in Table~\ref{tab:GELATO_PARAMS}. These are classified into two groups: Narrow and Balmer.
Emission-line \textit{species} within their respective \textit{groups} can optionally have their redshifts and velocity dispersions ($\sigma$) tied together. Following \cite{Hviding2022}, both $z$ and $\sigma$ of {\it species} in the `Balmer' group are tied. Flux ratio constraints based on atomic physics are naturally enforced for all emission lines within a given species. For \oiii\, and Balmer lines,  we also included the possibility of additional components designated as `Outflow' and `Broad', respectively. Such components, often associated with AGNs, do not share the $z$ and $\sigma$ values of their narrow counterparts.
To avoid fitting physically  unrealistic components, the line centroids ($v_0$) based on the redshifts ($z_0$) and dispersions ($\sigma$) of the Gaussian components were constrained to GELATO's predefined ranges\footnote{For the narrow components, the limits are 60~\kms\,$<\sigma <$ 500~\kms\ and $v_0$ within $\pm$~300~\kms\ from $z_0$. For the `Broad' and `Outflow' components, these restrictions are 1200~\kms\,$<\sigma <$ 6500~\kms\ with $v_0$ within $\pm600$~\kms, and  100~\kms\,$<\sigma <$ 750~\kms\ and $v_0$ within [-750, +150]\,\kms, respectively.
}.  These constraints are based on the emission line properties of large samples of quasars and galaxies from \citet{Mullaney13}, \citet{Shen11} and \citet{Hao2005}. 
%

To model the underlying continuum emission, GELATO uses a Trust Region Reflective \citep[TRF;][]{Branch1999} algorithm with a least-squares loss function to iteratively estimate the best-fitting parameters from an initial estimate of the model. 
After  masking out a 10,000~\kms\, region around each specified spectral line, the continuum is modeled using simple stellar populations (SSPs) spanning a range of ages and metallicities from the Extended MILES stellar library \citep[E-MILES;][]{Vazdekis2016}. To characterize the power-law emission originating from an unobscured AGN accretion disc, GELATO assesses whether the inclusion of a power-law component to the model leads to a substantial improvement in the fit. This validation is performed using an F-test, affirming significance if the improvement is $>3\sigma$ ($\sim 99.87\%$ confidence level).

The GELATO fits revealed that an additional `Broad' component to model the H$\alpha$ emission is required for 643/1908 systems. In 123 of these cases, the dispersion of the `Broad' component is close to the upper fitting bound (i.e.\ $\sigma \approx 6500$~km\,s$^{-1}$) and characterized by low S/N ($<$5 for 117/123).  Typically, a S/N$>$5 is used to identify robust `Broad' components \cite[][]{Escott2025}.  Adopting this threshold, we rejected `Broad' components in these 117 sources and refitted them using only `Narrow' components.
The distribution of the S/N of \oiii$\lambda5007$ from the final fits is shown in the last panel of Fig.~\ref{fig:fig1}. Since detection of this line is paramount for classifying sources into LERGs/HERGs \citep[e.g.,][]{Tadhunter1998, Buttiglione10, Best12, Pracy16}, we discarded 375/1908 sources where the S/N(\oiii\,)$<1$.
Further details on the fitted line parameters and examples of fits for the remaining 1533 radio sources are provided in the following sections.

\subsection{Separating AGN and SF galaxies} 
\label{sec:SF_AGN}

\begin{figure}[t]
\centerline{\vbox{
\centerline{\hbox{ 
\includegraphics[
trim = {0cm 0cm 0cm 0cm}, clip=true,
width=0.50\textwidth,angle=0]{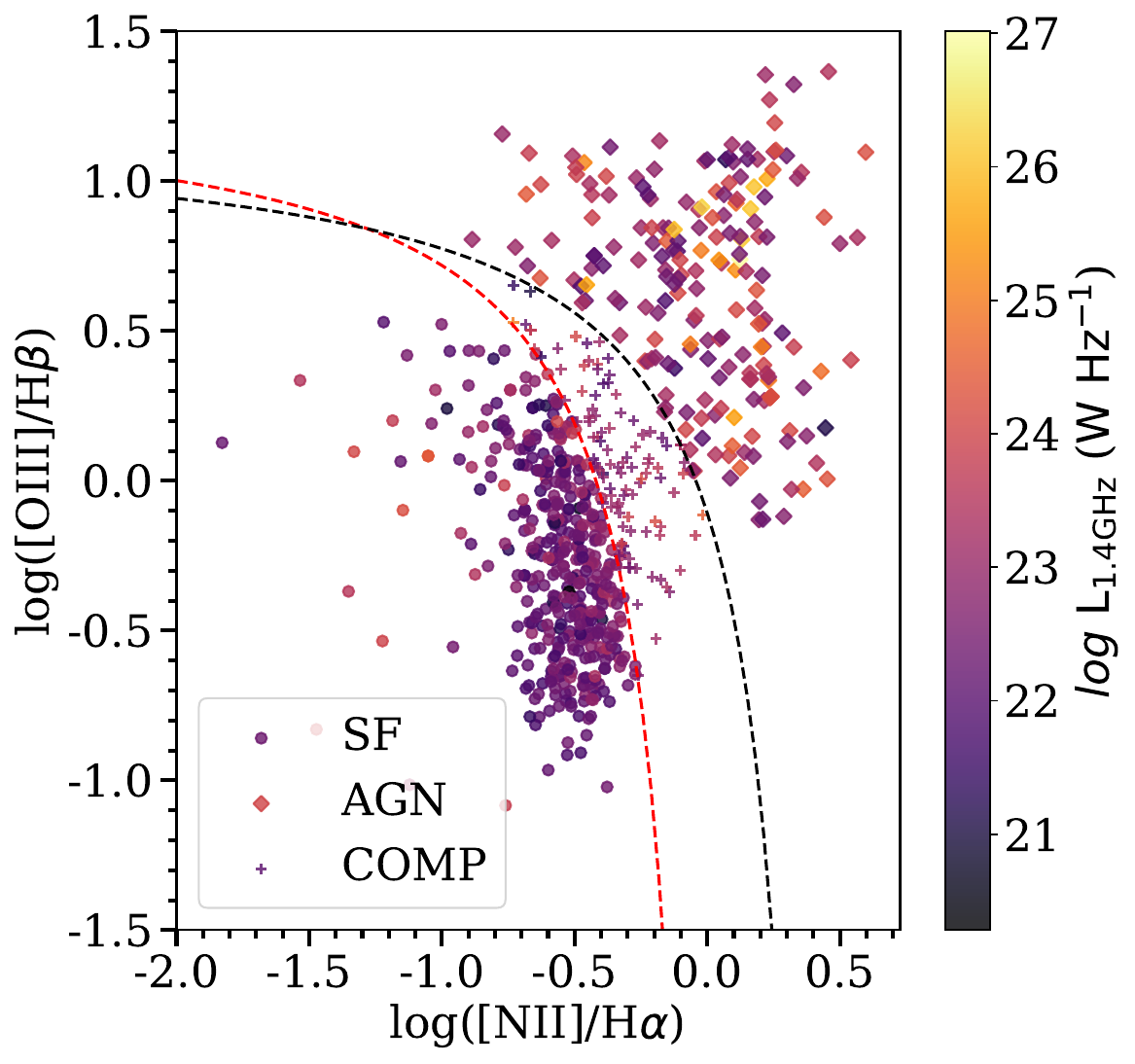}  
}} 
}}  
\vskip+0.0cm   
\caption{BPT1 classified AGN, SF, and Composite (COMP) galaxies color-coded according to their 1.4~GHz radio luminosity (L$_{1.4\,\rm GHz}).$
}
\label{fig:bpt1}
\end{figure}

We classified radio sources with the \oiii$\lambda5007$ line detected at S/N~$>1$ into either AGN or SF galaxies, predominantly utilizing the BPT diagnostic \citep[][]{Baldwin1981}. Fig.~\ref{fig:snr} presents the S/N distributions of various emission lines relevant for the classification. Based on the emission-line ratios alone (see Table~\ref{tab:class} for a summary), 730 sources were classified into AGN (196), SF (426) or composite (108) galaxies. Except for three cases, the remaining are based on BPT1: \oiii/H$\beta$ vs. \nii/H$\alpha$ (see Fig.~\ref{fig:bpt1}). BPT2 and BPT3 refer to the diagnostic diagrams of \oiii/H$\beta$ versus \sii/H$\alpha$ and \oi/H$\alpha$, respectively. These diagnostics classified one and two sources into SF and AGN categories, respectively.

\begin{table}[htbp]
    \centering
    \caption{Summary of classification of 1300 out of 1533 sources with \oiii\, detected at S/N$>$1 into AGN, SF or composite galaxies. 233 sources remain unclassified. Table~\ref{tab:lerg_herg_samp} presents the full sample.}
    \label{tab:SF_AGN}
    \begin{tabular}{cccc}
        \hline\hline
        Classification method & SF & AGN & Composite \\
        \hline
        BPT1 & 425 & 194$^\dag$ & 108 \\
        BPT2 & 1   & 0   & 0\\
        BPT3 & 0   & 2   & 0\\
        L$_{1.4\,\rm GHz}$ vs. L$_{\rm H\alpha}$ & 85 & 261 & -- \\
        $\rm L_{1.4GHz} > 10^{24}$\,W\,Hz$^{-1}$ & -- & 224 & --\\
        \hline\hline
    \end{tabular}
    \label{tab:class}
    \tablefoot{$^\dag$ Includes 68 radio-quiet AGN  with only narrow Balmer lines.}
\end{table}

\begin{figure}[t]
\centerline{\vbox{
\centerline{\hbox{ 
\includegraphics[
trim = {0cm 0cm 0cm 0cm}, clip=true,
width=0.50\textwidth,angle=0]{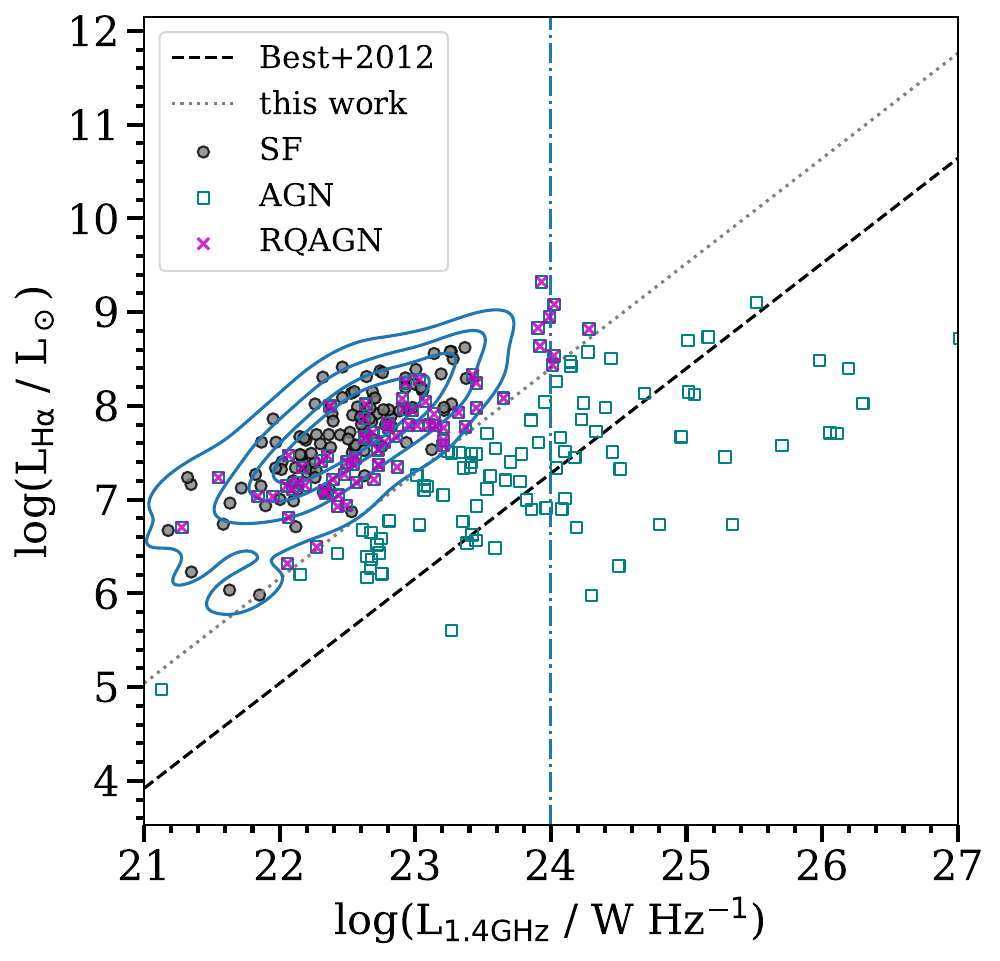}  
}} 
}}  
\vskip+0.0cm   
\caption{BPT classified AGN and SF galaxies  with only narrow Balmer lines in the L$_{\rm H\alpha}$ -- L$_{1.4\,\rm GHz}$ plane. The dashed line (black) marks the conservative division between AGN and SF galaxies from \citet[][]{Best12}, and the contours correspond to SF galaxies.  The dotted line represents the criteria used to exclude radio-quiet optical AGNs (RQAGN).  The vertical line marks the SF/AGN division, where sources with $\rm L_{1.4\,\mathrm{GHz}} > 10^{24}\,\mathrm{W\,Hz^{-1}}$ are classified as AGNs.}
\label{fig:lhalpha}
\end{figure}

In the remaining cases, not all four emission lines required for the BPT classification were reliably detected (S/N~$>3$) or covered within the wavelength range of the SDSS spectra. To classify these sources, we employ the L$_{\rm H\alpha}$ versus L$_{1.4\,\rm GHz}$ diagnostic plane. Both H$\alpha$ and 1.4\,GHz radio continuum luminosities serve as independent measures of the star-formation rate in SF galaxies, creating a well-defined correlation. Conversely, for radio-loud AGNs, a significant excess of radio emission over that estimated from the H$\alpha$ luminosity is expected. To formally separate these populations, \cite{Best12} utilized a conservative division line defined by $\log(\rm L_{H\alpha}/L_\odot) = 1.12\times (\log L_{1.4\,\rm GHz} (W\,Hz^{-1}) - 17.5)$. Fig.~\ref{fig:lhalpha} depicts our BPT-classified AGN and SF galaxies on this plane. Interestingly, a significant fraction of BPT classified AGNs lie in the SF region in this diagram, implying that in these sources the radio emission is predominantly due to star-formation rather than AGN activity. This, however, does not apply to `Broad'-lined objects, where the H$\alpha$ luminosity mainly originates from the AGN Broad Line Region (BLR) rather than star formation. Consequently, we restrict this planar analysis to systems with only narrow Balmer lines (Fig.~\ref{fig:lhalpha}). The distribution of SF galaxies, shown as contours in Fig.~\ref{fig:lhalpha}, suggests that an alternative division line (dotted), defined by $\log\rm (L_{H\alpha}/L_\odot) = 1.12\times (\log L_{1.4\rm GHz} (W\,Hz^{-1}) - 16.5)$, encapsulates the SF population more realistically. Using this modified division line, we first excluded 68 narrow-lined radio-quiet optical AGNs from the sample. Applying this same criterion to the aforementioned sources lacking full BPT coverage, we classified 346 sources into 261 AGNs and 85 SF galaxies. 

Of the remaining 457/1533 that could not be classified using any of the above methods, we classified 224 with luminosities exceeding L$_{1.4\,\rm GHz} > 10^{24}$\,W\,Hz$^{-1}$ as AGNs, since such high luminosities would require unrealistically large star-formation rates to be powered by star formation alone \citep[e.g.,][]{Pracy16, Murphy2011}. The bottom-right panel of Fig.~\ref{fig:snr} shows the location of sources in the L$_{1.4\,\rm GHz}$--$z$ plane based on the classification method. Evidently, most sources at higher redshifts are classified using the L$_{1.4\,\rm GHz}$ criterion.
%
Finally, 233 sources remain unclassified. We exclude these from further analysis. In summary, 1300 out of 1533 sources were classified into one of the following categories: AGN (681), SF (511), or composite (108) galaxies (Table~\ref{tab:class}). The full MALS-SDSS catalog is provided in Table~\ref{tab:lerg_herg_samp}.
%

\subsection{AGN classification into LERGs/HERGs} 
\label{sec:LERG_HERG}

Here we classify the 613 radio-loud AGNs, obtained after excluding 68 radio-quiet sources from the 681 AGNs identified in the previous section, into LERGs and HERGs using the following set of diagnostics from \citep[][]{Best12}. 
%
\begin{enumerate}

    \item A source is a HERG (LERG) if all six emission lines, \oiii\,, \nii\,, \sii\,, \oi\,, H$\alpha$, and H$\beta$ are detected, and EI $>$ 0.95 (EI $<$ 0.95) by at least 1$\sigma$. Typical HERG and LERG spectra classified using this method ({\tt Diagnostic 1}) are shown in Fig.~\ref{fig:EI}. 

    \item If only four out of six above-mentioned lines are detected,  classification is determined at  $>$1$\sigma$ significance using the diagnostics ({\tt Diagnostic 2}) of \cite{Kewley06} (see Fig.~\ref{fig:K06} for representative spectra).

    \item Classification is based on the rest equivalent width (REW) of \oiii\, $>$ 5\AA\ (HERG) or $<$ 5\AA\ (LERG) at 1$\sigma$ significance ({\tt Diagnostic 3}; see Fig.~\ref{fig:OIIIEW} for representative spectra).

    \item (1) $-$ (3) are repeated, but without  applying the $1\sigma$ criterion ({\tt Diagnostic 4}).

    \item For weak line galaxies with \oiii\,, \nii\, and H$\alpha$ detections at $>1\sigma$, we apply the diagnostics of \cite{CidFernandes10} based on \oiii\,/H$\alpha$ and \nii\,/H$\alpha$ ratios ({\tt Diagnostic 5}; see Fig.~\ref{fig:CF10} for representative spectra).

\end{enumerate}
%
Usually, these five diagnostics are applied in a preferential order to classify  narrow-line AGNs \citep[e.g.,][]{Buttiglione10, Best12}. 
 Rather than applying the five diagnostics sequentially, as in \citet[][]{Best12}, we apply all diagnostics simultaneously and combine their output using a weighted framework to obtain a more robust classification. To achieve this, we introduce a confidence measure ($C$) and a reliability factor ($R$) to quantify data quality and the intrinsic strength of each diagnostic.    
The final significance ($S$)  of a source's classification is then the weighted average of all available diagnostics, $S$ = $\frac{\sum_{i=1}^5 R_i\cdot C_i}{\sum_{i=1}^5 R_i}$.  Note that we assign a negative sign for LERGs and a positive sign for HERGs. This framework assigns a continuous, interpretable significance score, where, $|S| \in [0,1]$.

The details of calculations of $C$ and $R$ are provided in Appendix~\ref{sec:optical_fit}.  We evaluated various weighting schemes to calculate $S$, ultimately adopting $R_1 = 1$, $R_2 = R_3 = 0.9$, $R_4 = 0.8$, and $R_5 = 0.5$ for AGNs fitted using only narrow components.  For the 82 AGNs requiring an additional broad H$\alpha$ component with integrated S/N~$>5$, the criterion of EI = 0.95 ({\tt Diagnostic 1}) is inapplicable.  For these we set $R_1$ = 0.  From the narrow-only and narrow+broad subsets, we identified  58 HERGs / 318 LERGs and 41 HERGs / 9 LERGs, respectively.  The prevalence of HERGs (82\%) among AGNs with broad Balmer components is consistent with the findings of \citet{Buttiglione10}.
In summary, starting with  613 AGNs with 531 (narrow-only) + 82 (narrow+broad) AGN, we classified a total of  426 AGNs into  99 HERGs and 327 LERGs. The remaining  187 sources could not be classified due to a lack of emission lines satisfying the S/N criteria.
Note that the initial subset of 613 AGNs still included 82 sources with a broad H$\alpha$ component. For these the AGN activity cannot be unambiguously confirmed but the contamination is too small to affect the statistical results presented here.

For consistency check, we cross-matched our MALS-SDSS sample of  426 HERGs/LERGs with \citet[][]{Best12} and found 43 common sources. Due to the choice of weights $R_i$, we find an excellent agreement,  as only one source shows classification mismatch: J144017.88$+$055630.1, which is a LERG in \cite{Best12}, but a HERG ($S = 0.42$) in our weighting-based classification scheme.
Additionally,  41 sources are common with \citet[][]{Pracy16}.  There are three mismatches: J094319.09$-$000423.7, J145322.99$+$045802.6, and J015517.11$+$002741.4.  In our scheme, J094319.09$-$000423.7 is a borderline LERG (EI = 0.85$\pm$0.04; $S$ $\sim$ $-0.33$). However, its spectrum shows a strong \oiii\,$\lambda5007$ line (REW = 89.9$\pm$2.3~\AA) and a weak 4000~\AA\, break, indicating HERG-like features. Similarly, J145322.99$+$045802.6 is also a borderline LERG (EI = 0.89$\pm$0.05; $S$ $\sim$ $-0.01$) with a strong \oiii\,$\lambda5007$ line (REW = 54.0$\pm$3.9~\AA) and a weak 4000~\AA\, break.  J015517.11$+$002741.4 is a HERG in our classification scheme with EI = 1.11$\pm$0.07 and $S = 0.59$, while it is a LERG in \citet{Pracy16}.
We do not reclassify these marginal mismatches.

\subsection{Properties of HERGs and LERGs}

\begin{figure*}[]
\centerline{\vbox{
\centerline{\hbox{ 
\includegraphics[trim = {0cm 0cm 0cm 0cm}, clip=true,
width=\textwidth,angle=0]{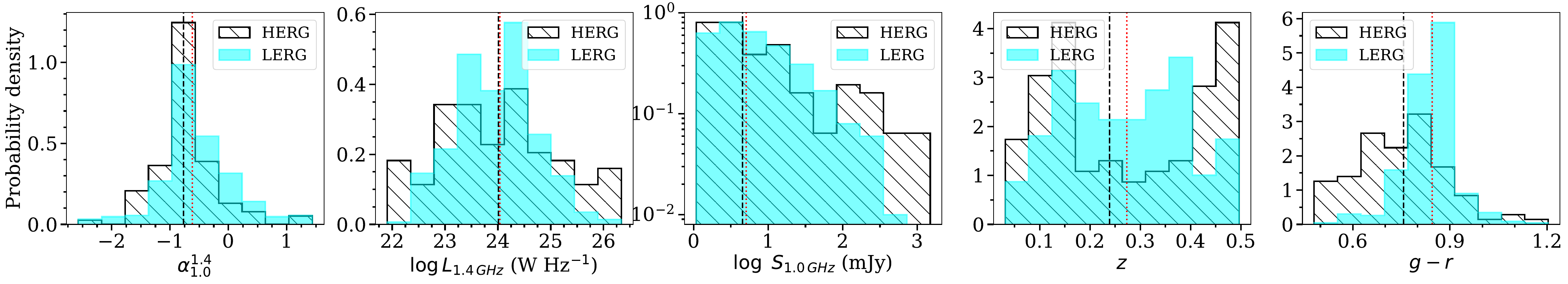}  
}} 
}}  
\vskip+0.0cm   
\caption{Distribution of $\alpha_{1.0}^{1.4}$ ({\it first}), L$_{1.4\rm GHz}$ ({\it second}), S$_{1.0\rm GHz}$ ({\it third}), $z$ ({\it fourth}), and  $k$-corrected $g-r$ ({\it fifth}) for the LERG and HERG samples.  The red and black dashed lines mark the median values for LERGs and HERGs, respectively.}
\label{fig:LERG_HERG_compare}
\end{figure*}

\begin{table}
\begin{center}
\centering
\tabcolsep=2pt
\footnotesize
\caption{Statistical comparison between LERG (327) and HERG (99) samples. The reported uncertainties correspond to the median absolute deviation (MAD).}
\begin{tabular}{lccc}
\hline\hline
    Parameter                         &  median LERG  & median HERG  &  KS test $p$-value  \\
\hline
$\alpha_{1.0}^{1.4}$                  &   $-0.62\pm 0.29$    &   $-0.77\pm 0.20$    &   0.002 \\
$\log \rm L_{1.4\rm GHz}$ (W\,Hz$^{-1}$)  &  $24.04\pm 0.51$     & $24.01\pm 0.81$      & 0.04 \\
$S_{1.0\rm GHz}$ (mJy)                &  $5.07\pm 3.33$      &  $4.53\pm 2.94$      &  0.3 \\
$z$                                   & $ 0.27\pm 0.10$      & $0.24\pm 0.15$       &  0.007 \\ 
$g-r$ ($k-$corrected)                 & $0.85\pm 0.04$       &  $0.76\pm 0.09$      &   $3.3\times 10^{-15}$ \\     
\hline\hline
\end{tabular}
\label{tab:LERG_HERG_comp}
\end{center}
\end{table}

Fig.~\ref{fig:LERG_HERG_compare} compares HERGs and LERGs in terms of spectral indices ($\alpha_{1.0}^{1.4}$), radio luminosity (L$_{1.4\rm GHz}$), flux density at 1.0~GHz (S$_{1.0\rm GHz}$), redshift, and $k-$corrected\footnote{Following the methodology of \cite{Chilingarian2010}.} optical color ($g-r$).  
The statistics and KS-test $p$-values in Table~\ref{tab:LERG_HERG_comp} show that LERGs and HERGs differ by $\gtrsim 3\sigma$ in $\alpha_{1.0}^{1.4}$ and optical colors, with no significant difference in $S_{1.0\rm GHz}$. Notably, while LERGs show a slightly higher median $\log \rm L_{1.4\rm GHz}$, HERGs span the full luminosity range (Fig.~\ref{fig:LERG_HERG_compare}, {\it second} panel). The two populations also differ at $>2\sigma$ in redshift, with LERGs distributed at higher median $z$.
 These differences in $\rm L_{1.4\rm GHz}$ and $z$ are primarily driven by the selection effects discussed in Section~\ref{sec:SF_AGN}. The only astrophysically meaningful differences are observed in spectral index and optical $g - r$ color.
The slightly steeper median $\alpha_{1.0}^{1.4}$ for HERGs likely reflects their association with FRII-type, lobe-dominated radio galaxies,  whereas LERGs comprise both FRI and FRII types \citep{Buttiglione10, Best12, Heckman14}. 
A more pronounced difference ($>3\sigma$) is seen in $k-$corrected optical $g - r$ color. LERGs are significantly redder, reflecting older, more massive, and less star-forming host galaxies. 
This confirms the finding of \citet[][]{Best12} that LERGs have redder optical colors than HERGs of the same radio luminosity, here extended to a broader luminosity range (log\,L$_{\rm 1.4GHz}$\,(W\,Hz$^{-1}$) $\sim$ $21.1-27.0$) and higher redshifts ($z<0.5$).

\begin{figure*}[]
\centerline{\vbox{
\centerline{\hbox{ 
\includegraphics[trim = {0cm 0cm 0cm 0cm}, clip=true,
width=\textwidth,angle=0]{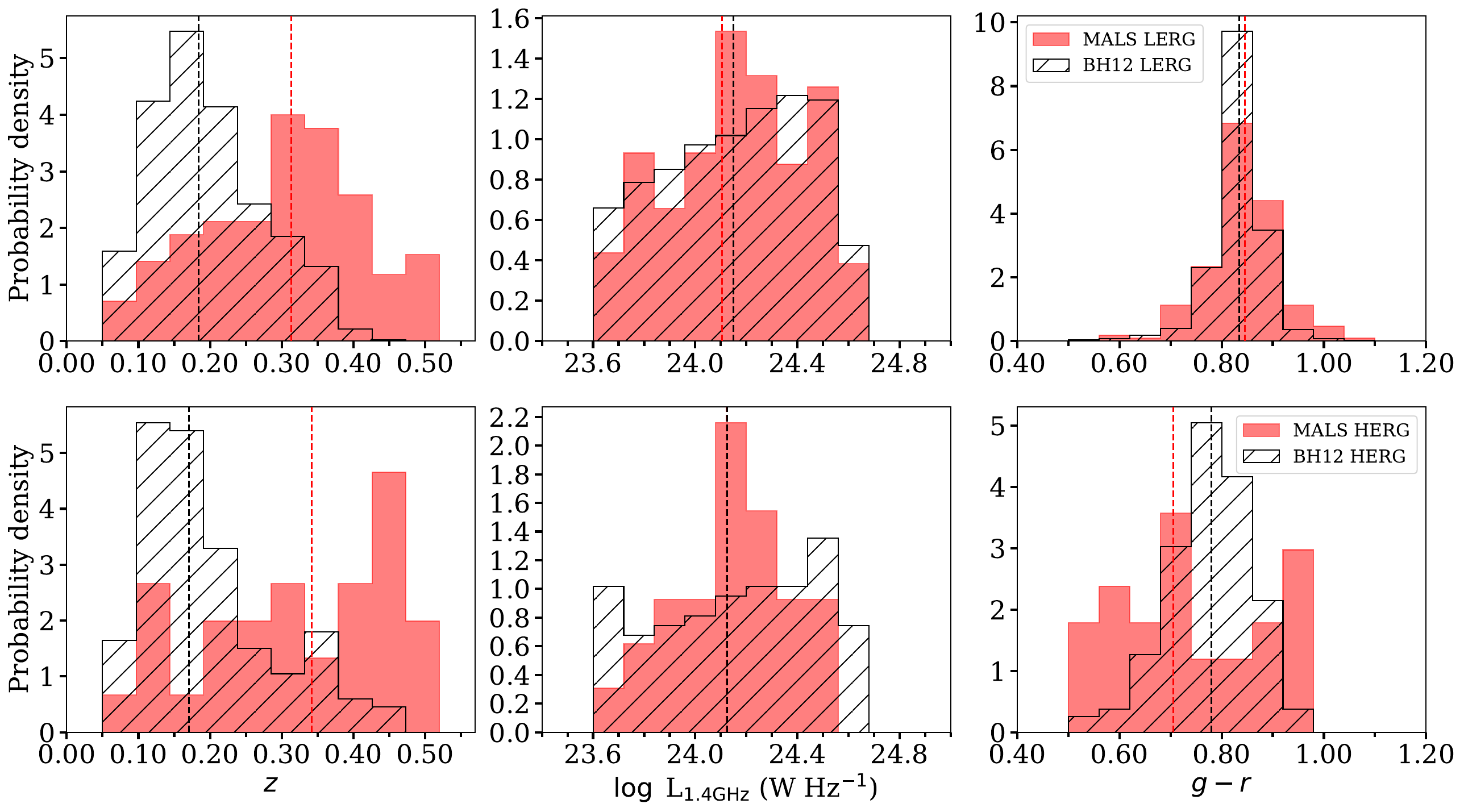}  
}} 
}}  
\vskip+0.0cm   
\caption{Histograms comparing the distributions of $z$, $\rm L_{1.4\,GHz}$, and $k$-corrected ($g-r$) for LERGs ({\it top panel}) and HERGs ({\it bottom panel}), using the MALS and \cite{Best12} samples matched in L$_{\rm 1.4\,GHz}$. 
}
\label{fig:match_BH12_MALS}
\end{figure*}

\begin{figure}[t]
\centerline{\vbox{
\centerline{\hbox{ 
\includegraphics[
trim = {0cm 0cm 0cm 0cm}, clip=true,
width=0.45\textwidth,angle=0]{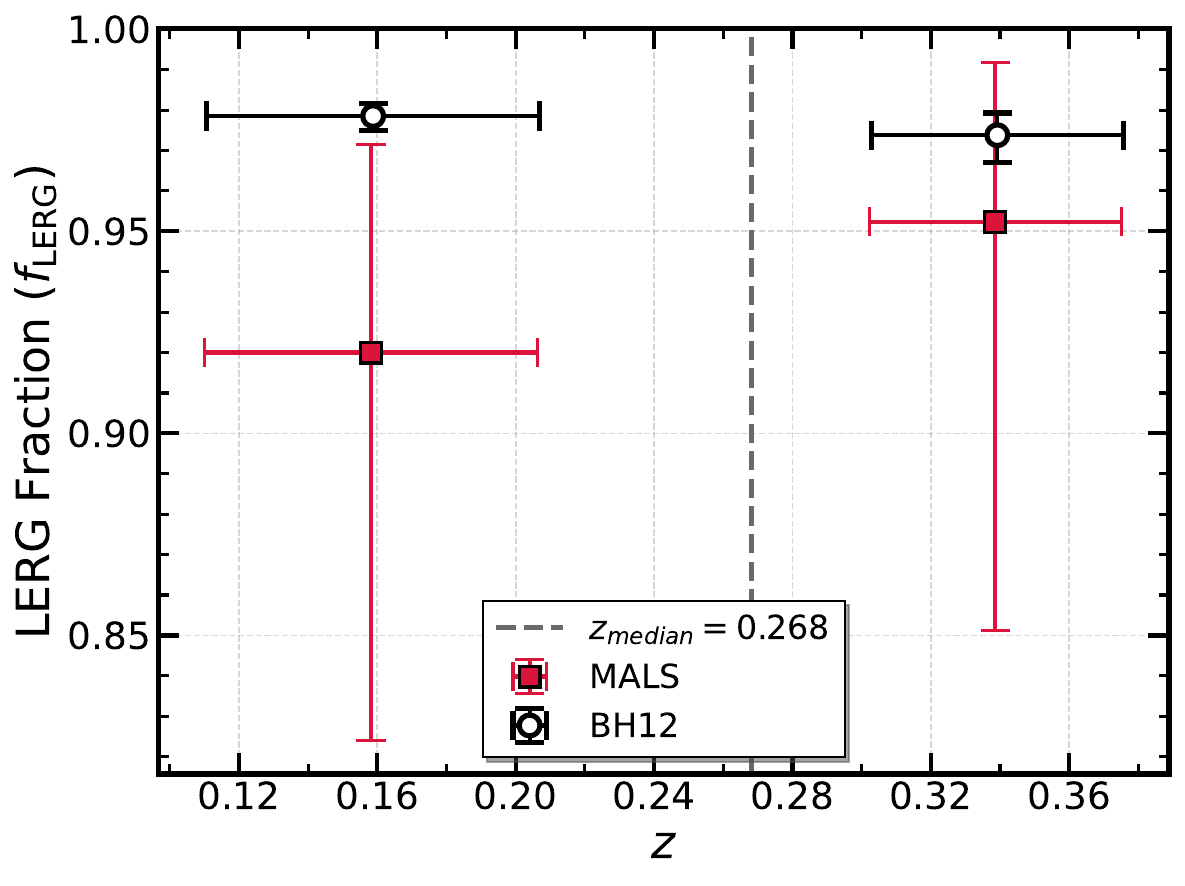}  
}} 
}}  
\vskip+0.0cm   
\caption{Comparison of LERG fractions between the BH12 \citep{Best12} and MALS samples, evaluated within two redshift bins. Horizontal error bars denote the width of each redshift bin, while vertical error bars represent the binomial uncertainties on the derived fractions.}

\label{fig:LERG_fraction}
\end{figure}

 In Fig.~\ref{fig:match_BH12_MALS} we show the distribution of redshift, 1.4~GHz radio luminosity, and $k$-corrected $g-r$ color for $\rm L_{1.4GHz}$-matched MALS and \citet{Best12} samples. The $k$-corrected $g-r$ colors are consistent for the LERGs at $p=0.05$ level, while the marginal $<3\sigma$ ($p=0.003$) discrepancy for HERGs is likely an artifact of small-number statistics in the MALS subset ($N=32$).
 To compare the LERG fractions between the MALS and \citet{Best12} samples objectively, we matched the datasets in both redshift and $\rm L_{1.4,GHz}$. The samples were partitioned at the median redshift of the MALS sample, and the low- and high-$z$ bins were analyzed independently.
From Fig.~\ref{fig:LERG_fraction}, in the low-$z$ bin, the  The LERG fractions are $92.0^{+5.2}_{-9.6}$\% for MALS and $97.9 \pm 0.3$\% for BH12 (Binomial uncertainties), representing only a $\sim 1\sigma$ difference.  In the high-$z$ bin, the fractions are $95.2^{+3.9}_{-10.1}$\% (MALS) and $97.4 \pm 0.6$\% (BH12), which are again statistically consistent. To conclude, we find no statistically significant evidence for a redshift dependence in the LERG fraction between the two matched samples.

\section{\hi\ 21-cm absorption search}    
\label{sec:21cm}  
%
\begin{figure*}[]
    \begin{center}
    \vbox{
        \hbox{
            \includegraphics[width=0.25\textwidth]{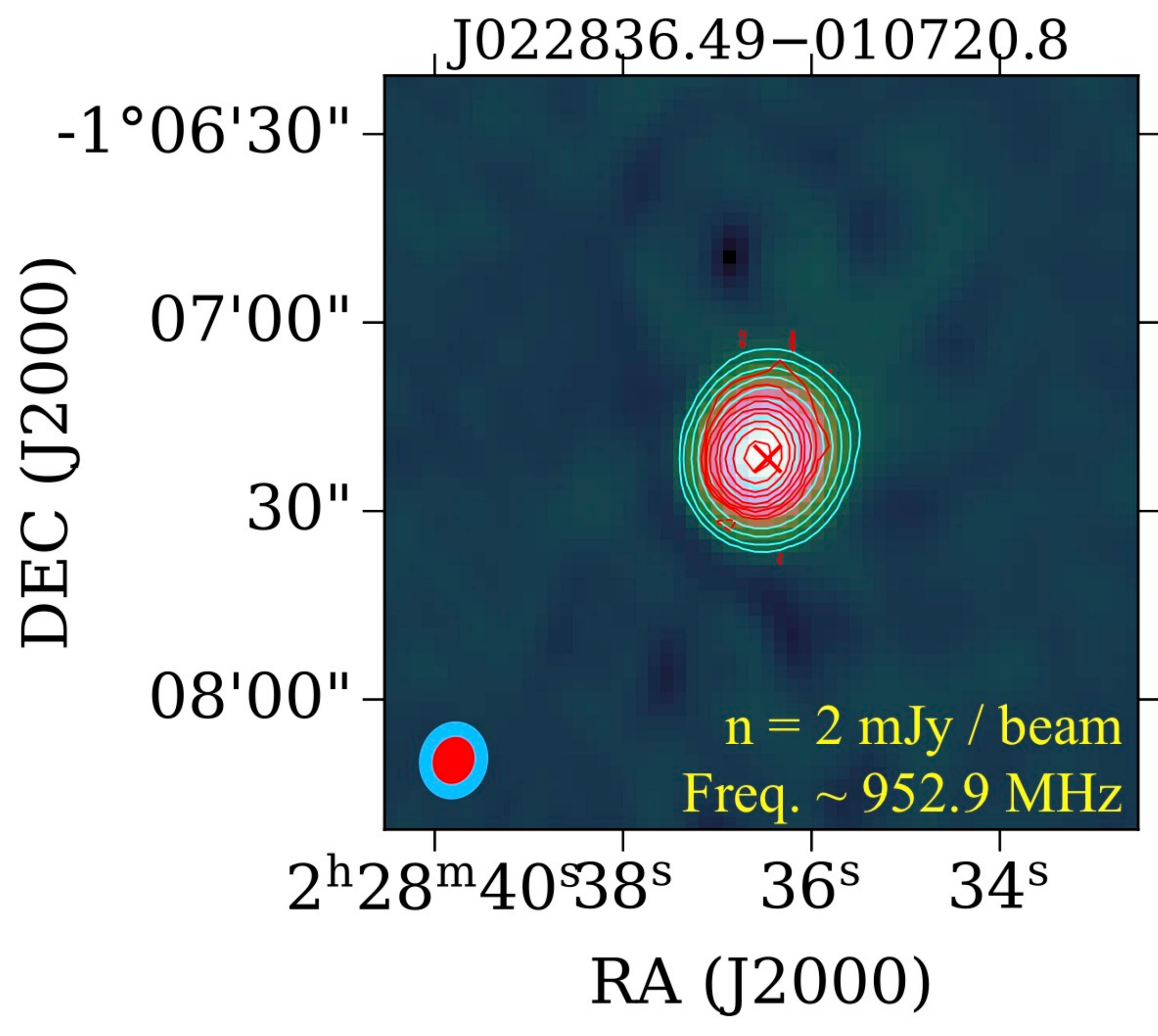}
            \includegraphics[width=0.57\textwidth]{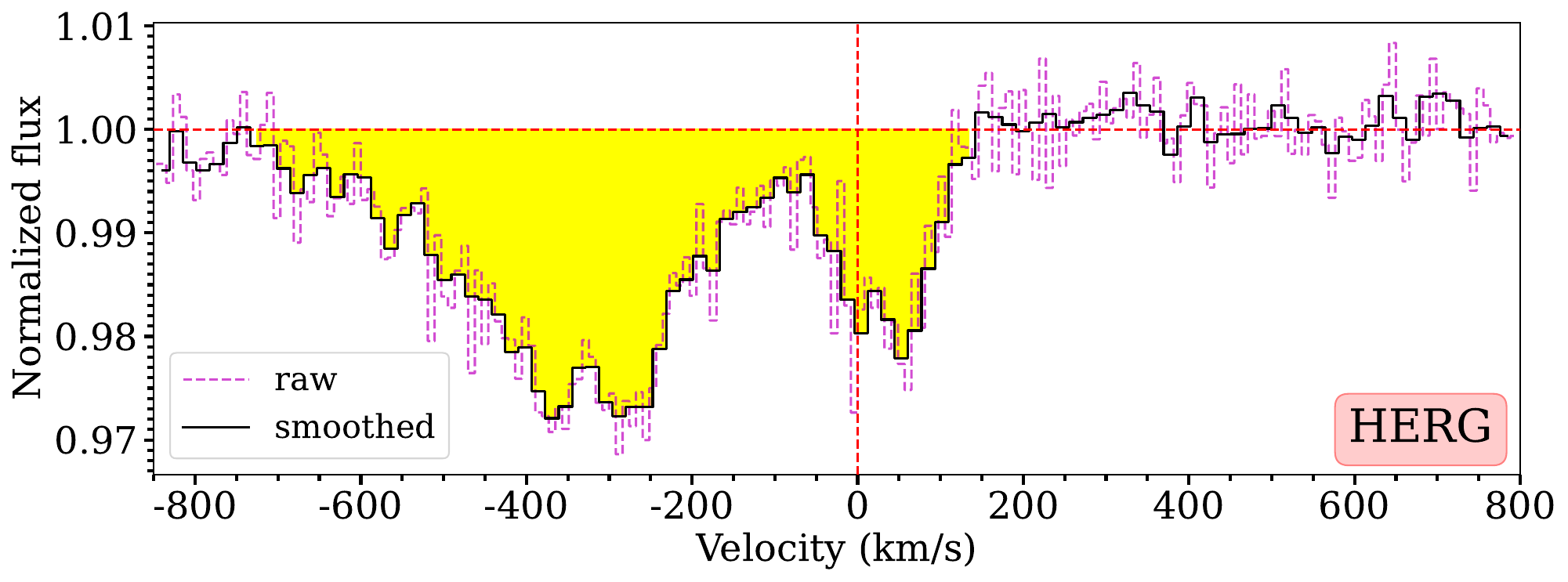}
            }
        \hbox{
            \includegraphics[width=0.25\textwidth]{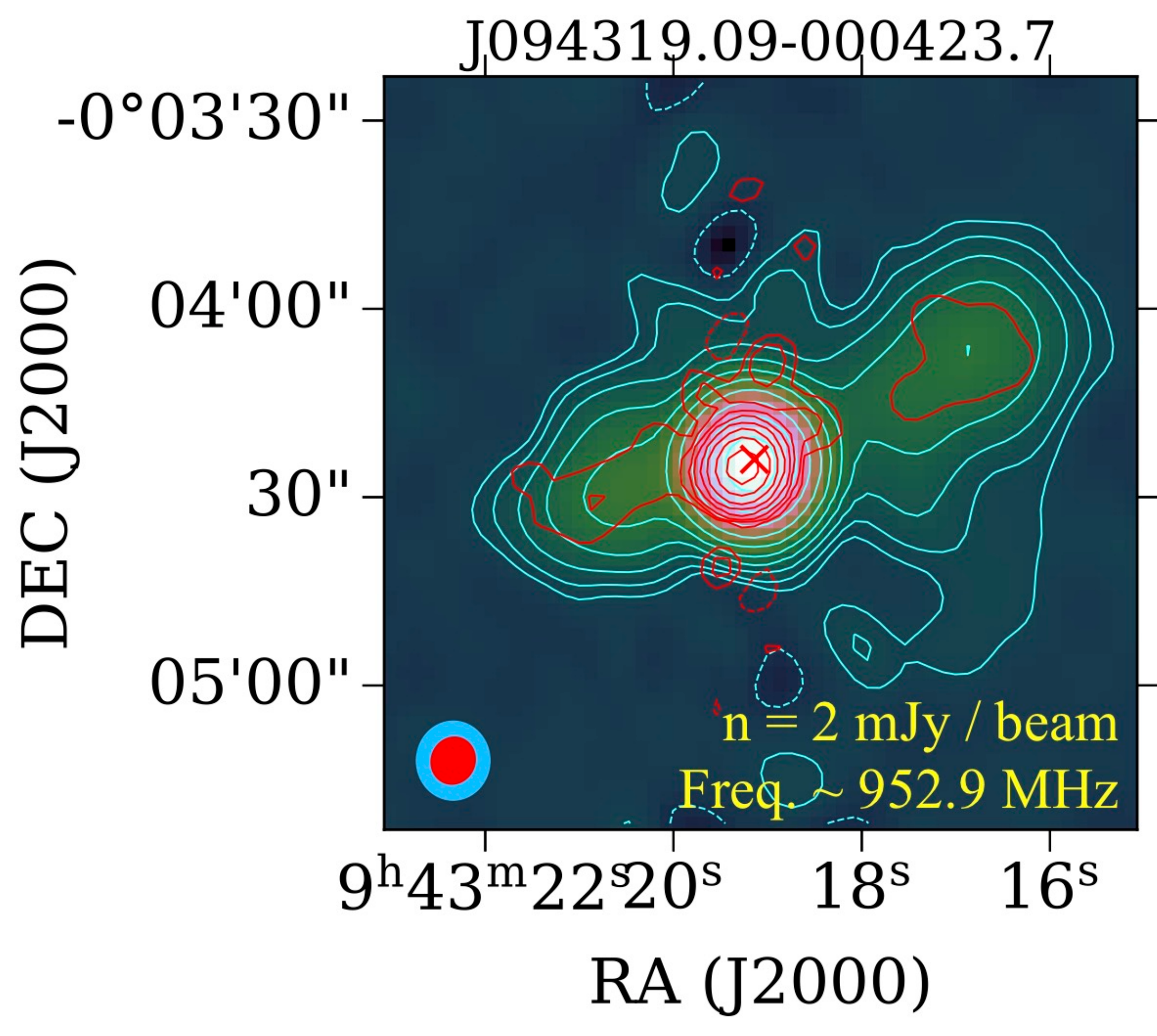}
            \includegraphics[width=0.57\textwidth]{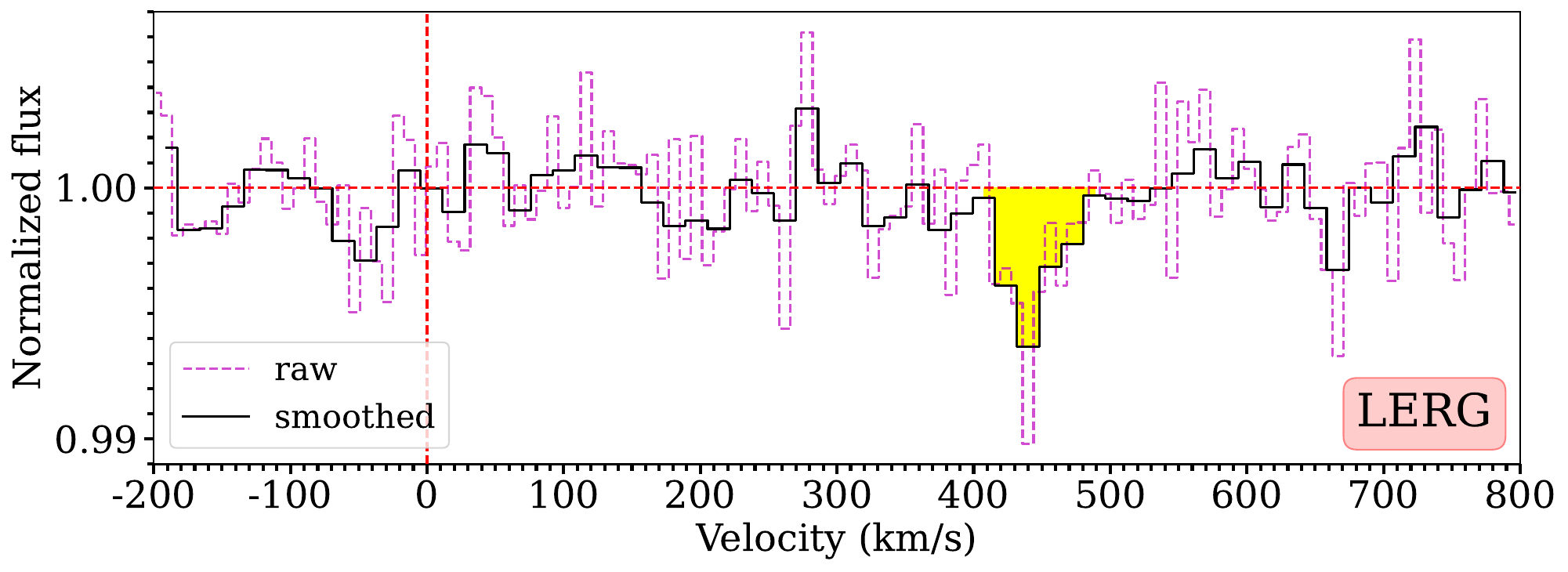}
        }
        \hbox{
            \includegraphics[width=0.25\textwidth]{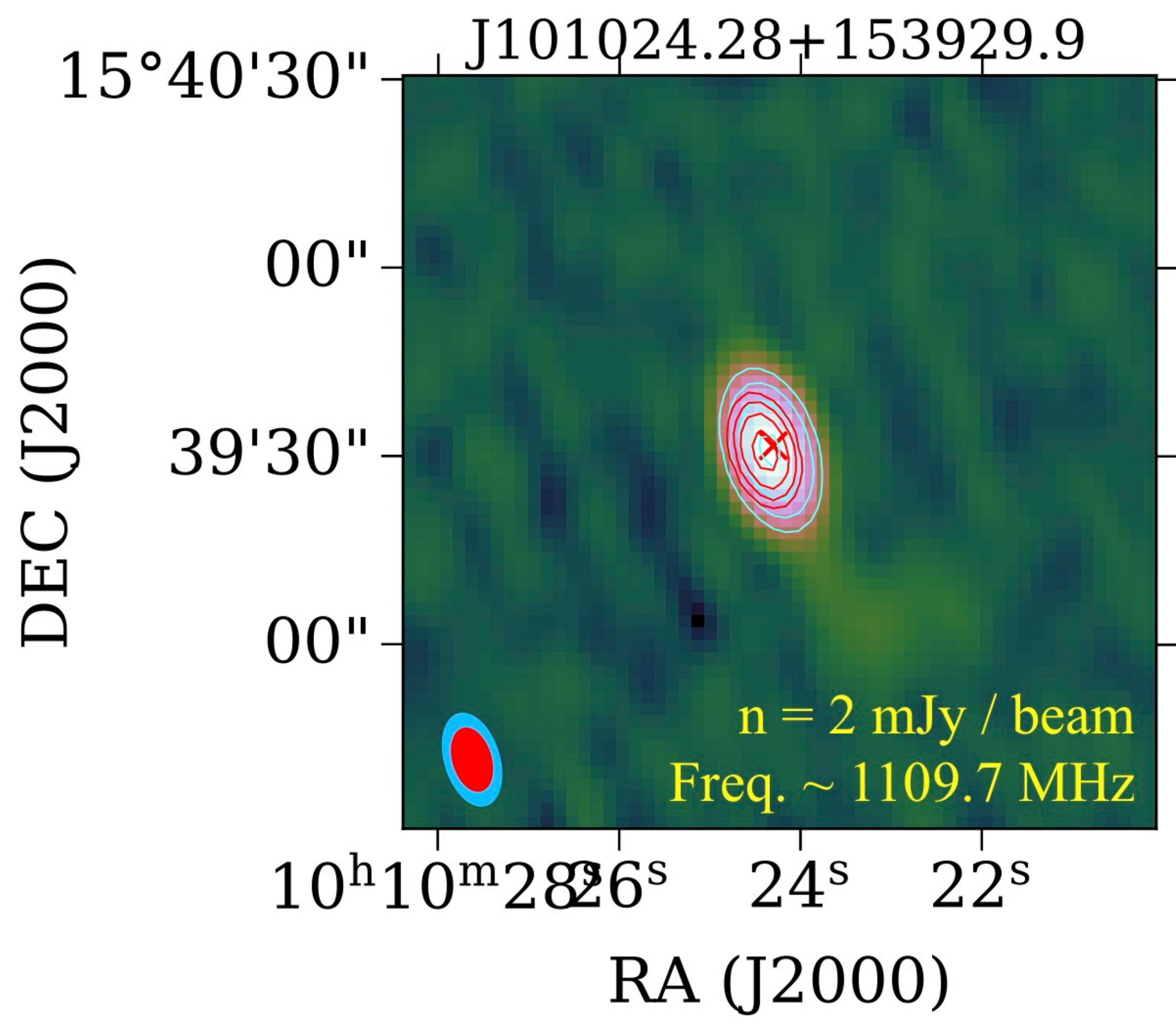}
            \includegraphics[width=0.57\textwidth]{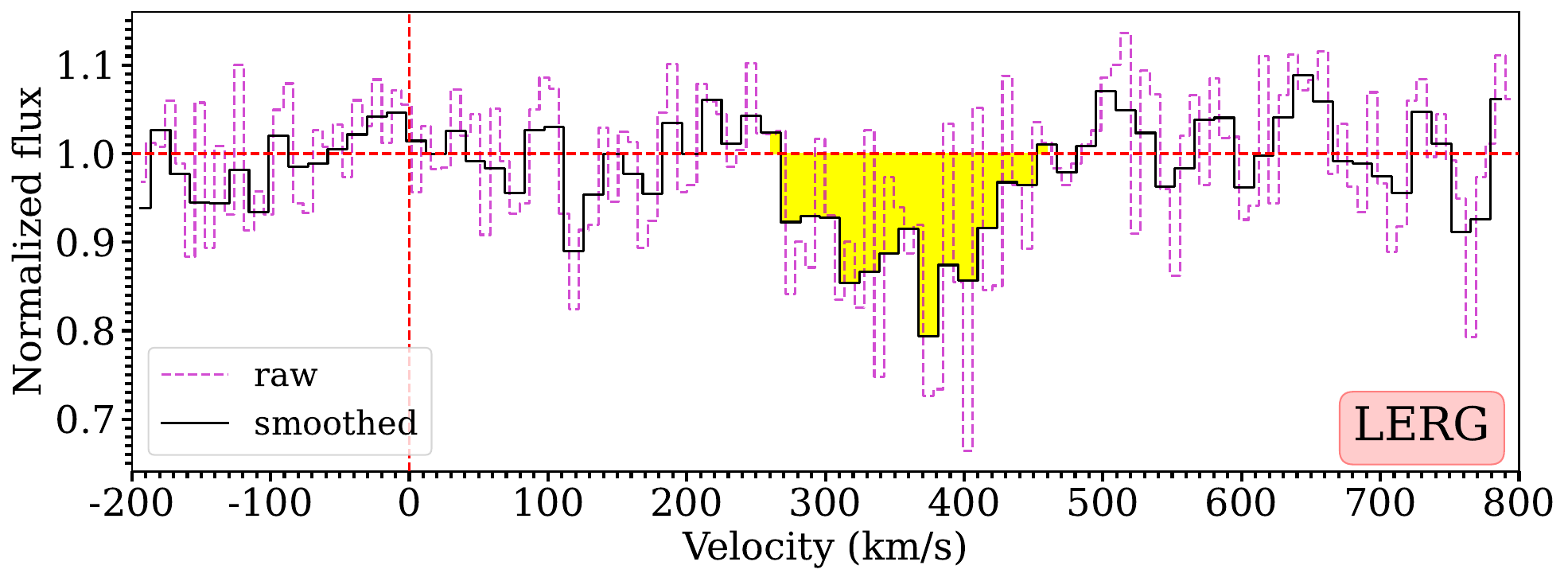}
        }   
        \hbox{
            \includegraphics[width=0.25\textwidth]{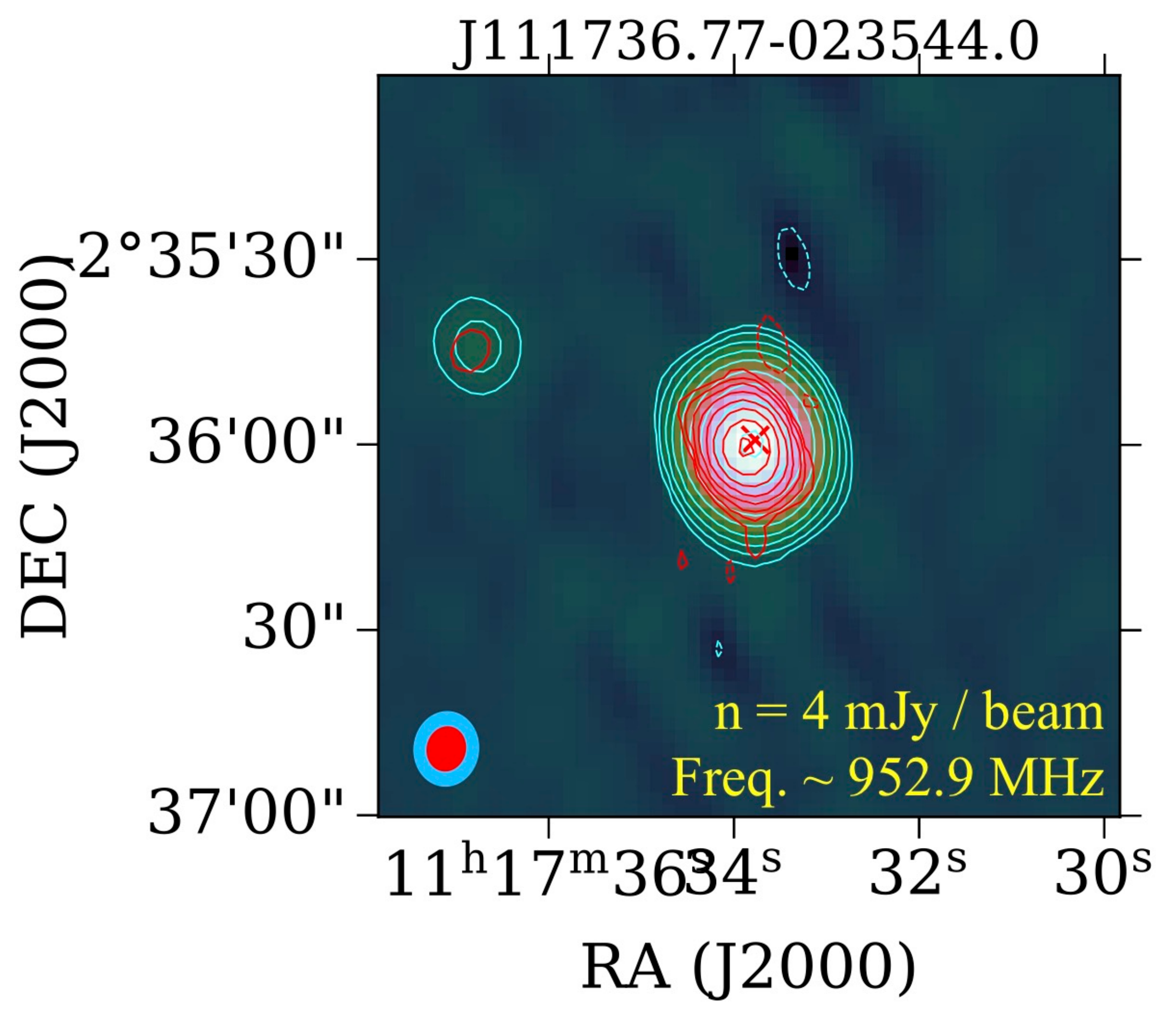}
            \includegraphics[width=0.57\textwidth]{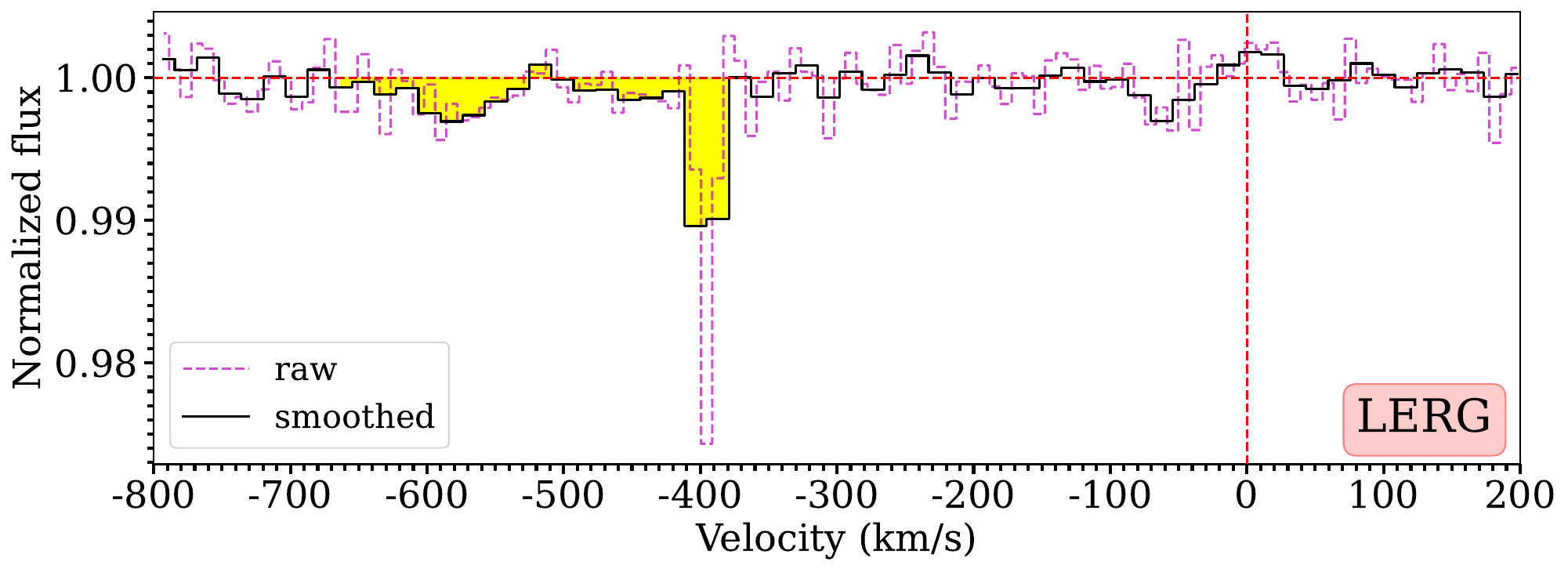}
        }
        \hbox{
            \includegraphics[width=0.25\textwidth]{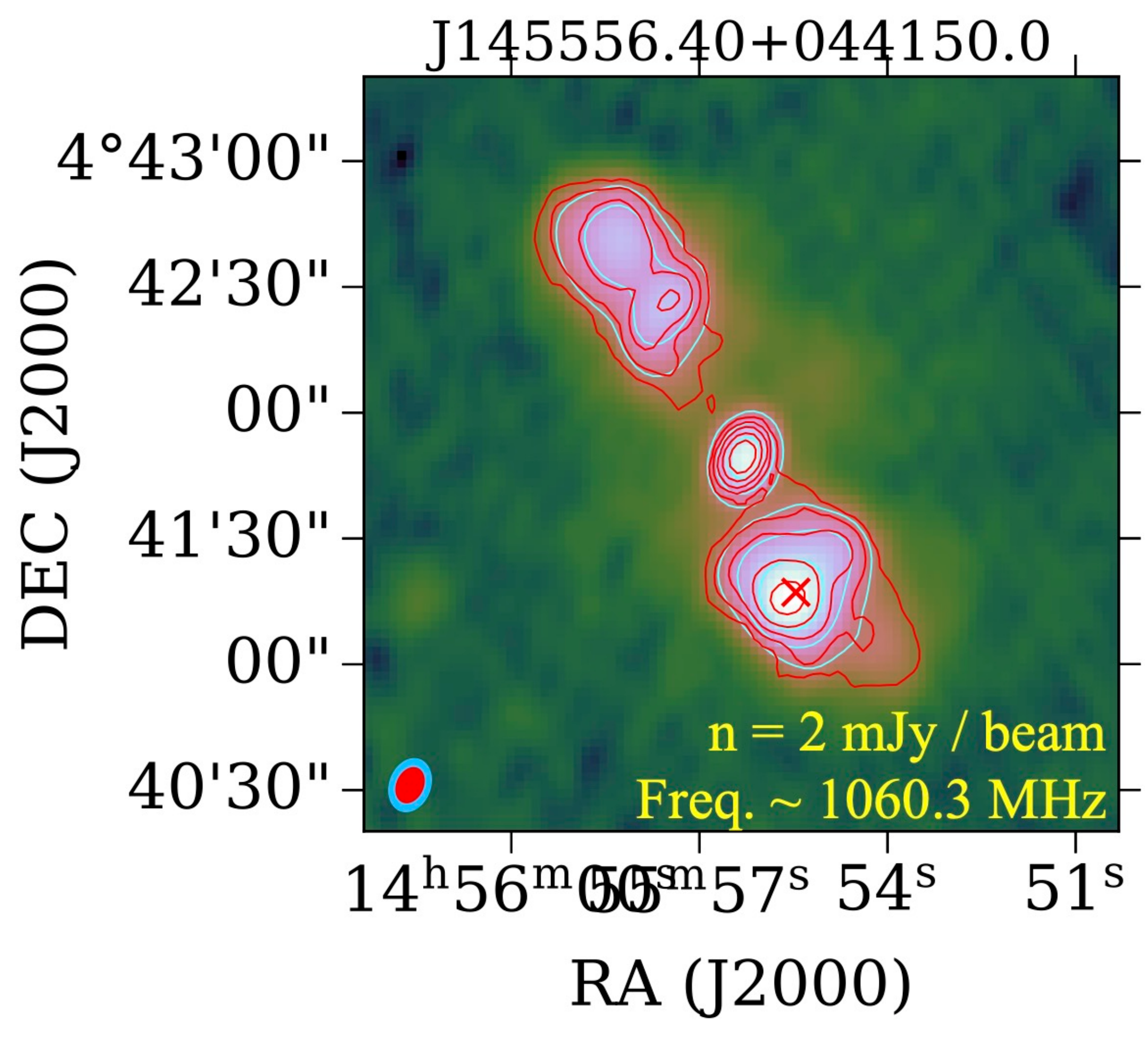}
            \includegraphics[width=0.57\textwidth]{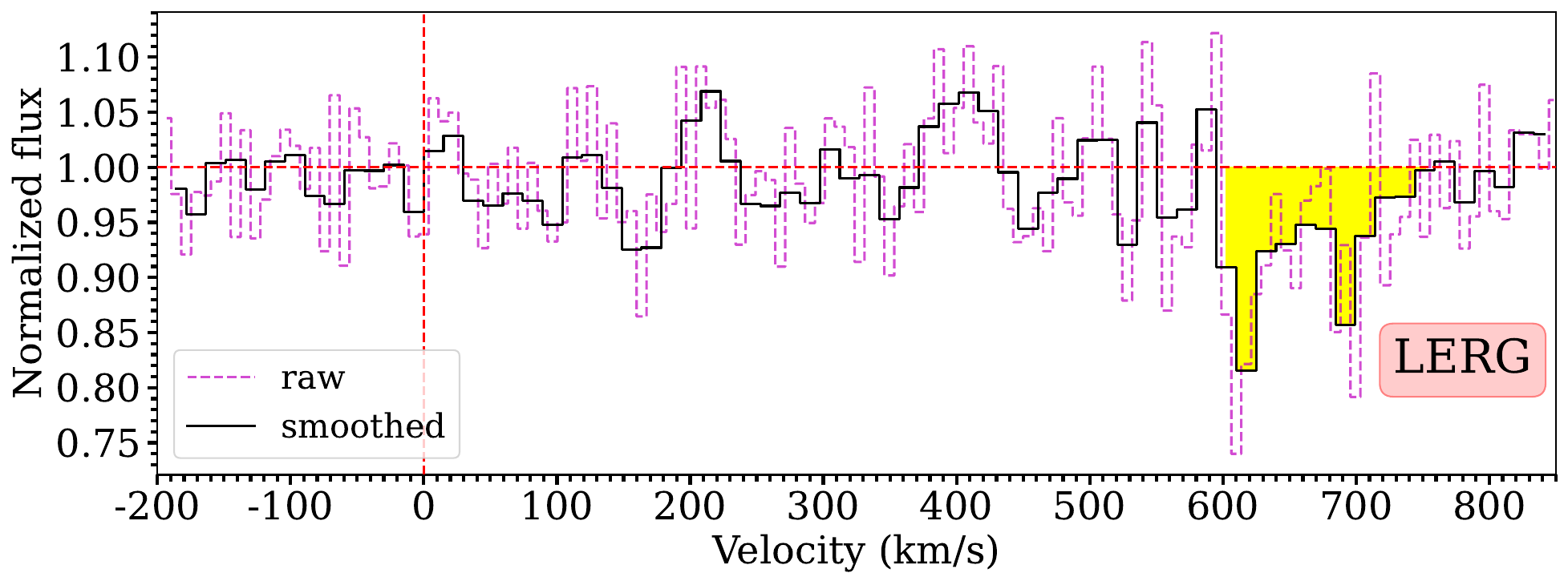}
        }
        }
    \end{center}
    \caption{The radio continuum images (left) and normalized absorption profiles (right) for the five sources listed in Table~\ref{tab:detections_2mJy}.  The zero velocity corresponds to the optical redshift ($z_{em}$). Cyan contour levels at $n \times$ [-1, 1, 2, 4, 8, 16, ...]) represent the background continuum near the redshifted 21-cm line frequency.   The corresponding $n$ values and the image frequency are provided within the respective panels. 
    Red contours denote the highest-resolution MALS L-band images (1643.0~MHz), with $n$ values of 0.5, 5, 1.5, 4, and 0.3~mJy\,beam$^{-1}$, respectively. Cyan and red filled ellipses (\textit{bottom left}) indicate the restoring beams for the background and high-resolution images, respectively. Spectra were smoothed with a 3-pixel Boxcar and decimated by a factor of two. Shaded velocity ranges correspond to the integrated optical depth ($\int\tau\,dv$) values listed in Table~\ref{tab:detections_2mJy}.
    }
   \label{fig:21cmdet}
\end{figure*}

\begin{table*}[]
\setlength{\tabcolsep}{4pt}
\caption{Details of \hi\ 21-cm absorption detections. \label{tab:detections_2mJy} } 
\begin{tabular}{lccccccccc}
\hline\hline
MALS ID  & $z_{em}$(opt)  & $z_{abs}$ & $\Delta v$ & $\sigma_{rms}$ & FWBN & S$_{\rm norm}$ & $\int \tau\ dv$ &  N(\hi\,) & Class \\
   &     &   & (\kms\,)  & (mJy\,beam$^{-1}$) & (\kms\,) & (mJy) & (\kms\,) & (cm$^{-2}$) & \\
(1) & (2) & (3) & (4) & (5) & (6) & (7) & (8) & (9) & (10) \\ 
\hline
J022836.49$-$010720.8 & 0.47115 & 0.46977 & 8.1 & 0.8 & 880 & 231.4 &  12.0$\pm$0.3 & (22.2$\pm$0.3)$\times 10^{20}$ & HERG \\ 
J094319.09$-$000423.7 & 0.46435 & 0.46650 & 8.1 & 1.9 & 122  & 734.5 & 0.27$\pm$0.07 & (49$\pm$9)$\times 10^{18}$ & LERG  \\ 
J101024.28$+$153929.9 & 0.28620 & 0.28781 & 7.1 & 0.9 & 214  & 14.1  & 21$\pm$2      & (37$\pm$3)$\times 10^{20}$ & LERG  \\ 
J111736.77$-$023544.0 & 0.47290 & 0.47096 & 8.1 & 1.2 & 324  & 757.3 & 0.62$\pm$0.08 & (11.2$\pm$0.8)$\times 10^{19}$ & LERG \\ 
J145556.40$+$044150.0 & 0.34714 & 0.34988 & 7.4 & 0.8 & 220  & 17.2  & 13$\pm$1      & (24$\pm$2)$\times 10^{20}$ & LERG \\  
\hline
\hline
\end{tabular}
\textcolor{black}{
\small{
    Column\,1: MALS-SPW2 source ID. Column\,2: Optical spectroscopic redshift from SDSS. Column\,3: Redshift corresponding to absorption peak. Column\,4: Velocity resolution at the redshifted 21-cm line frequency. Column\,5: Unsmoothed spectral rms. Column\,6: Full width between nulls (FWBN) of the absorption profile. Column\,7: Flux density used to normalize the spectrum. Column\,8: Velocity integrated optical depth. Column\,9: Column density estimated assuming $T_s = 100$~K and $f_c = 1$. Column\,10: Optical spectroscopic classification. }  \\
}
\end{table*}


 We selected 99 radio sources ($\rm S_{\rm 1.4\,GHz}>$ 4\,mJy)  with usable spectra from the MALS-SDSS sample at $z_{em}<0.5$ for an \hi\ 21-cm absorption line search.  This subset comprises 79 LERGs and  20 HERGs.  Of these, 22 were classified as AGN using BPT1,  36 using the $\rm L_{H\alpha}$ - $\rm L_{\rm 1.4GHz}$ criteria and 41 based on $\rm L_{\rm 1.4GHz}>10^{24}$ W\,Hz$^{-1}$.
The sample spans nearly six decades in radio luminosity and reaching flux densities an order of magnitude fainter than any prior targeted \hi\ survey. 
Following the absorption line search methodology of \citet[][]{Gupta25galhi}, we detected five absorbers. The resulting absorption profiles presented in Fig.~\ref{fig:21cmdet} exhibit widths spanning $\sim$100 to $\sim$1000\,\kms\, (see Table~\ref{tab:detections_2mJy} for details).  Notably, three absorbers are entirely redshifted with respect to $z_{em}$, likely representing infalling gas  or gas perturbed by merger activity.
Below, we first discuss the properties of detections in detail and then the statistical properties of the  overall sample, including \hi\ absorption non-detections.

\subsection{Properties of detections}
\label{sec:absprop}

\subsubsection{Kinematics of the absorbing gas}
\label{sec:abskin}

\begin{figure*}[]
\centerline{\vbox{
\centerline{\hbox{ 
\includegraphics[
trim = {0cm 0cm 0cm 0cm}, clip=true,
width=\textwidth,angle=0]{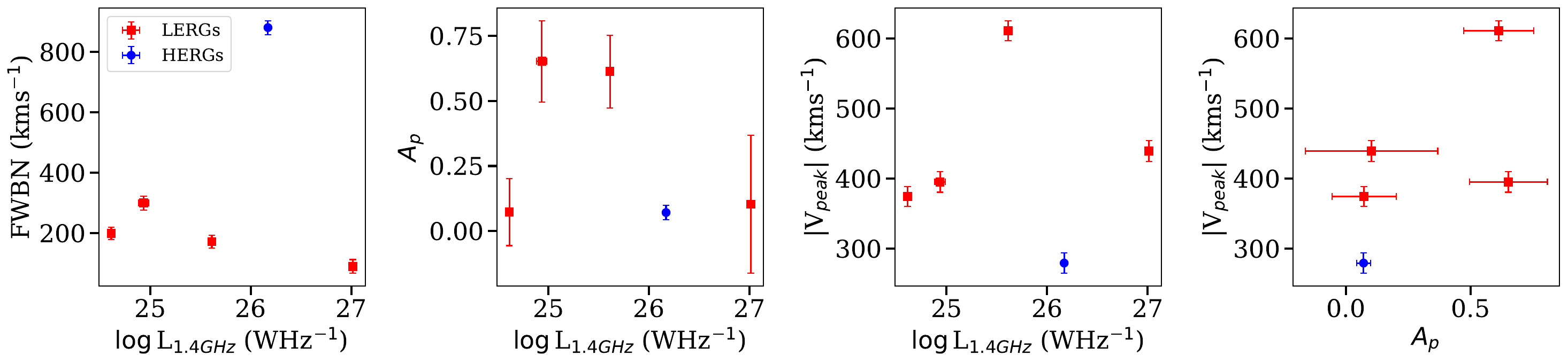}  
}} 
}}  
\vskip+0.0cm   
\caption{Distributions of absorption line properties for the five detections.  FWBN denotes the full width between nulls (FWBN) in \kms, $A_p$ measures the asymmetry of the line profile and $|V_{peak}|$ represents the shift of the absorption peak from $z_{em}$. The median uncertainty in $\rm\log L_{1.4GHz}$ is 0.003\,dex, and is not visible at the scale of the figure.  
}
\label{fig:P_Ap_Vnll_Vpeak}
\end{figure*}

 To characterize the gas kinematics, we use velocity shift of the absorption peak ($V_{\rm peak}$) relative to $z_{em}$, the full width between nulls (FWBN)  indicated by the shaded velocity range in Fig.~\ref{fig:21cmdet}, and the velocity-integrated optical depth ($\int \tau\, dv$). Asymmetries in the absorption profiles are quantified using the parameter:
\begin{equation}
    A_p = \frac{\Big |(\int\tau\,dv\large )_{Blue} - (\int\tau\,dv\large )_{Red}\big|}{\large(\int\tau\,dv\large)_{Total}},
\end{equation}
where $(\int\tau\,dv\large )_{Blue}$ and $(\int\tau\,dv\large )_{Red}$ represent the velocity-integrated optical depths  of the profile blue-ward and red-ward of $V_{\rm peak}$, respectively. Normalization by the total integrated optical depth, $(\int\tau\,dv\large )_{Total}$, allows a uniform comparison across profiles with varying strengths.
The first three panels of Fig.~\ref{fig:P_Ap_Vnll_Vpeak} present the distributions of FWBN, $|V_{\rm peak}|$ and $A_p$ as a function of $\rm L_{1.4\rm GHz}$ over $10^{24-27}$\,W\,Hz$^{-1}$.  Kendall-tau correlation tests show no significant trends among these parameters. This is consistent with an earlier finding of \cite{Deka2024b} in a sample of 19 \hi\ 21-cm absorbers associated with radio galaxies ($\rm L_{1.4,\rm GHz} >10^{26}$\,W\,Hz$^{-1}$). Similarly, no significant correlation is observed between $V_{\rm peak}$ and $A_p$ (Fig.~\ref{fig:P_Ap_Vnll_Vpeak}, last panel). 
 Among LERGs, we find a wide range in $A_p$ (0.07--0.65) and significant velocity offsets from $z_{\rm em}$ ($>350$~\kms), indicative of kinematically disturbed absorbing gas, possibly driven by jet-ISM interaction. This is consistent with studies of more powerful (L$_{1.4\,\mathrm{GHz}} \sim 10^{23}$--$10^{26}$~W\,Hz$^{-1}$) and lower-redshift ($z<0.2$) HERG/LERG samples \citep[e.g.][]{Glowacki2017, Chandola2020}. Due to the single HERG detection in our sample, a robust kinematic comparison between the two classes is currently not feasible.

\subsubsection{Radio morphology}
\label{sec:absmorph}

Radio morphology may provide further insight into the origin of the observed \hi\ kinematics. Two notable examples are the LERGs J0943$-$0004 and J1455$+$0441.  In both cases, the radio emission is extended and the absorbing gas appears to be `infalling' relative to the systemic redshift. The MALS image of J0943$-$0004 displays a dominant `core' and diffuse lobes spanning $\sim$47 kpc,  with the absorption against the radio `core'.  This gas infall, as  suggested by the presence of double-peaked \oiii\,$\lambda\lambda 4959, 5007$ emission lines symmetrically displaced around the systemic redshift, may be merger-driven \citep[][]{Dutta19}.  Furthermore, Pan-STARRS images (Fig.~\ref{fig:opt_images}) also reveal a nearby galaxy at a projected separation of $\sim5^{\prime\prime}$ (projected separation $\sim 30$~kpc).
In contrast, the absorption towards J1455$+$0441 is detected against  a hotspot located asymmetrically near the AGN. This potentially represents gas associated with a  faint companion galaxy.   

For the remaining three cases, the radio emission is unresolved  at the MALS angular resolution.  However, at VLBI scales, J0228$-$0107 reveals two components at 4.4\,GHz but it is unclear whether these represent a core-jet structure or a pair of lobes.  J1010-1539 exhibits a core-jet morphology at 4.8\,GHz while J1117$-$0235 shows a complex morphology with multiple components  possessing steep spectral indices based on images at 4.8 and 2.3\,GHz.  
Both J0228$-$0107 and J1117$-$0235 show well-detached, blueshifted absorption components. In fact, the bulk of absorption in the five sources is completely blue or redshifted. Future VLBI spectroscopy and high-resolution molecular line observations will be essential to reveal the origin of the absorbing gas \citep[see][]{Combes24}.

\subsection{Statistical properties}
\label{sec:statprop}

The search within $\pm$2000\,\kms\ of the redshifted \hi\ 21-cm line frequency yielded five detections from the sample  of 99 radio loud AGNs (79 LERGs and  20 HERGs). This corresponds to a detection rate of $\sim$5\%, uncorrected for completeness or the varying optical depth sensitivity across the spectra.  The associated absorbers typically exhibit broad line widths of $\sim$100\,\kms, although, occasionally narrow (FWHM$\sim$20\,\kms) lines are also detected \citep[e.g.,][]{Gupta06, Maccagni17}.
Based on the completeness analysis in \citet[][their Fig.~7]{Gupta25galhi}, the completeness fraction for injected absorbers with $\sigma$ = 8\,\kms\ (FWHM $\sim$ 18.9\,\kms) at peak S/N $>6$ (integrated S/N $>8$) is near-unity. 
Notably, four of the five detections reported here have an integrated S/N $>$ 8, the exception is  J0943-0004 (integrated S/N = 4).
Adopting 20\,\kms\ as line width for non-detections, we estimate detection rates for a range of $3\sigma$ integrated optical depth cut-off limits over 0.1 - 10\,\kms\ (see Table~\ref{tab:detrates}).  
Clearly, the detection rates for HERGs and LERGs agree within uncertainties.

\begin{figure*}[]
\centerline{\vbox{
\centerline{\hbox{ 
\includegraphics[trim = {0cm 0cm 0cm 0cm}, clip=true,
width=0.8\textwidth,angle=0]{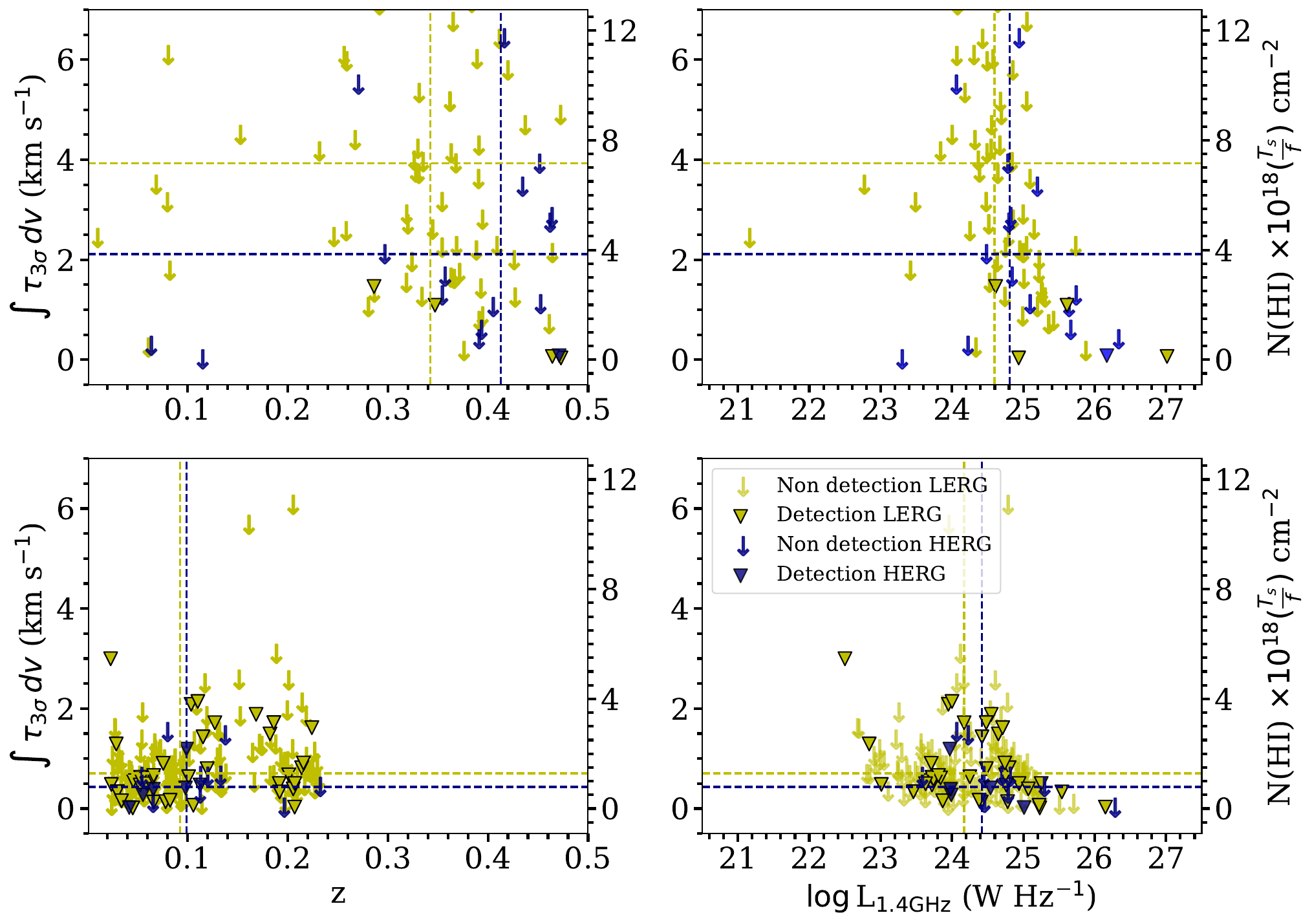}  
}} 
}}  
\vskip+0.0cm   
\caption{Top row: MALS sources; Bottom row: 185 (170 LERGs, 15 HERGs) sources from \cite{Maccagni17}. Legends are same across all panels. Vertical and horizontal dashed lines are median values. For better visibility 22 MALS sources (19 LERGs,  4 HERGs) are excluded from the panels with $\int\tau_{3\sigma}\,dv > 7$~\kms.   
}
\label{fig:lzmir}
\end{figure*}

Fig.~\ref{fig:lzmir} ({\it top row}) shows optical depth measurements versus 1.4\,GHz luminosity and redshift  for the MALS sample. Splitting the sample at the median, the detection rates for the cut-off $\int \tau_{3\sigma}\,dv < 10$~km\,s$^{-1}$ in the low/high bins are: L$_{1.4\,\rm GHz}$ -- $2.3^{+5.3}_{-1.9}\%$ (1/43) and $9.3^{+7.3}_{-4.4}\%$ (4/43); redshift -- $4.6^{+6.1}_{-3.0}\%$ (2/43) and $7.0^{+6.8}_{-3.8}\%$ (3/43). While consistent within uncertainties, notably there is a skew in detections towards high L$_{1.4\,\rm GHz}$. The high-L$_{1.4\,\rm GHz}$ bin is significantly ($>3\sigma$) more sensitive (median $\int\,\tau_{3\sigma}\,dv = 2.1$~km\,s$^{-1}$ vs.\ 4.2~km\,s$^{-1}$ in the low bin), and accounts for 4/5 detections. 

\begin{table*}[h]
\tabcolsep=2pt
\scriptsize
\centering
\caption{Detection rates in the MALS and \citet{Maccagni17} samples, with Poisson errors calculated following \citet{Gehrels86}. Values for absorbers with an integrated optical depth S/N $>$ 8 are provided in brackets \citep[details in][]{Gupta25galhi}. \label{tab:detrates}}
\begin{tabular}{l|ccc|ccc|ccc}
\hline \hline
Optical depth & Number of$^{\star}$ & Number of$^{\star\star}$ & Percentage & Number of$^{\star}$ & Number of$^{\star\star}$ & Percentage & Number of$^{\star}$ & Number of$^{\star\star}$ & Percentage\\
cutoff (\kms\,) & sight lines & Detections &  & sight lines & Detections & & sight lines & Detections & \\
\hline
    \multicolumn{10}{c}{} \\
    \multicolumn{10}{c}{\large MALS sample: 79 LERGs, 20 HERGs} \\
    \multicolumn{10}{c}{} \\
\hline
    & \multicolumn{3}{c|}{HERGs} & \multicolumn{3}{c|}{LERGs} & \multicolumn{3}{c}{combined} \\
\hline
0.1 & 2 & 1 (1) & $50^{+50}_{-41}$ ($50^{+50}_{-41}$) & 2 & 2 (0) & $100^{+0}_{-65}$ ($0^{+92}_{-0}$) & 4 & 3 (1) & $75^{+25}_{-41}$ ($25^{+58}_{-21}$) \\
0.2 & 2 & 1 (1) & $50^{+50}_{-41}$ ($50^{+50}_{-41}$) & 3 & 2 (0) & $67^{+33}_{-43}$ ($0^{+61}_{-0}$) & 5 & 3 (1) & $60^{+40}_{-33}$ ($20^{+46}_{-17}$)\\
0.3 & 3 & 1 (1) & $33^{+67}_{-28}$ ($33^{+67}_{-28}$) & 4 & 1 (0) & $25^{+58}_{-21}$ ($0^{+46}_{-0}$) & 7 & 2 (1) & $29^{+38}_{-19}$ ($14^{+33}_{-12}$)\\
0.5 & 4 & 1 (1) & $25^{+58}_{-21}$ ($25^{+58}_{-21}$) & 4 & 1 (0) & $25^{+58}_{-21}$ ($0^{+46}_{-0}$) & 8 & 2 (1) & $25^{+33}_{-16}$ ($13^{+29}_{-10}$)\\
1.0 & 5 & 1 (1) & $20^{+46}_{-17}$ ($20^{+46}_{-17}$) & 7 & 0 (0) & $0^{+26}_{-0}$ ($0^{+26}_{-0}$) & 12 & 1 (1) & $8^{+19}_{-7}$ ($8^{+19}_{-7}$)\\
5.0 & 14 & 1 (1) & $7^{+16}_{-6}$ ($7^{+16}_{-6}$) & 50 & 2 (2) & $4^{+5}_{-3}$ ($4^{+5}_{-3}$) & 64 & 3 (3) & $5^{+5}_{-3}$ ($5^{+5}_{-3}$)\\
10.0 & 18 & 1 (1) & $6^{+13}_{-5}$ ($6^{+13}_{-5}$) & 68 & 2 (2) & $3^{+4}_{-2}$ ($3^{+4}_{-2}$) & 86 & 3 (3) & $3^{+3}_{-2}$ ($3^{+3}_{-2}$)\\
\hline
    \multicolumn{10}{c}{} \\
    \multicolumn{10}{c}{\large \cite{Maccagni17} sample: 170 LERGs, 15 HERGs} \\
    \multicolumn{10}{c}{} \\
\hline
0.1 & 2 & 1 (1) & $50^{+50}_{-41}$ ($50^{+50}_{-41}$) & 7 & 3 (3) & $43^{+42}_{-23}$ ($43^{+42}_{-23}$) & 9 & 4 (4) & $44^{+35}_{-21}$ ($44^{+35}_{-21}$)\\
0.2 & 3 & 1 (1) & $33^{+67}_{-28}$ ($33^{+67}_{-28}$) & 14 & 7 (6) & $50^{+27}_{-18}$ ($43^{+26}_{-17}$) & 17 & 8 (7) & $47^{+23}_{-16}$ ($41^{+22}_{-15}$)\\
0.3 & 4 & 2 (2) & $50^{+50}_{-32}$ ($50^{+50}_{-32}$) & 21 & 7 (6) & $33^{+18}_{-12}$ ($29^{+17}_{-11}$) & 25 & 9 (8) & $36^{+16}_{-12}$ ($32^{+16}_{-11}$)\\
0.5 & 8 & 4 (4) & $50^{+40}_{-24}$ ($50^{+40}_{-24}$) & 54 & 13 (10) & $24^{+9}_{-7}$ ($19^{+8}_{-6}$) & 62 & 17 (14) & $27^{+8}_{-7}$ ($23^{+8}_{-6}$)\\
1.0 & 12 & 4 (4) & $33^{+26}_{-16}$ ($33^{+26}_{-16}$) & 117 & 25 (22) & $21^{+5}_{-4}$ ($19^{+5}_{-4}$) & 129 & 29 (26) & $23^{+5}_{-4}$ ($20^{+5}_{-4}$)\\
5.0 & 15 & 3 (3) & $20^{+20}_{-11}$ ($20^{+20}_{-11}$) & 168 & 25 (21) & $15^{+4}_{-3}$ ($13^{+3}_{-3}$) & 183 & 28 (24) & $15^{+4}_{-3}$ ($13^{+3}_{-3}$)\\
10.0 & 15 & 2 (2) & $13^{+18}_{-9}$ ($13^{+18}_{-9}$) & 170 & 13 (12) & $8^{+3}_{-2}$ ($7^{+3}_{-2}$) & 185 & 15 (14) & $8^{+3}_{-2}$ ($8^{+3}_{-2}$) \\
\hline\hline
\end{tabular}
\tablefoot{
\small{
    Both samples are scaled to a common velocity resolution of 20~\kms\ and integrated optical depth sensitivities are calculated assuming a Gaussian line with FWHM = 20~\kms. $^{\star}$ number of sight lines with $3\sigma$ integrated optical depth sensitivity better than optical depth cutoff; $^{\star\star}$ number of detection with $\int\tau\,dv >$ optical depth cutoff.}
}
\end{table*}

\begin{figure}[]
\centerline{\vbox{
\centerline{\hbox{ 
\includegraphics[trim = {0cm 0cm 0cm 0cm}, clip=true, width=0.45\textwidth,angle=0]{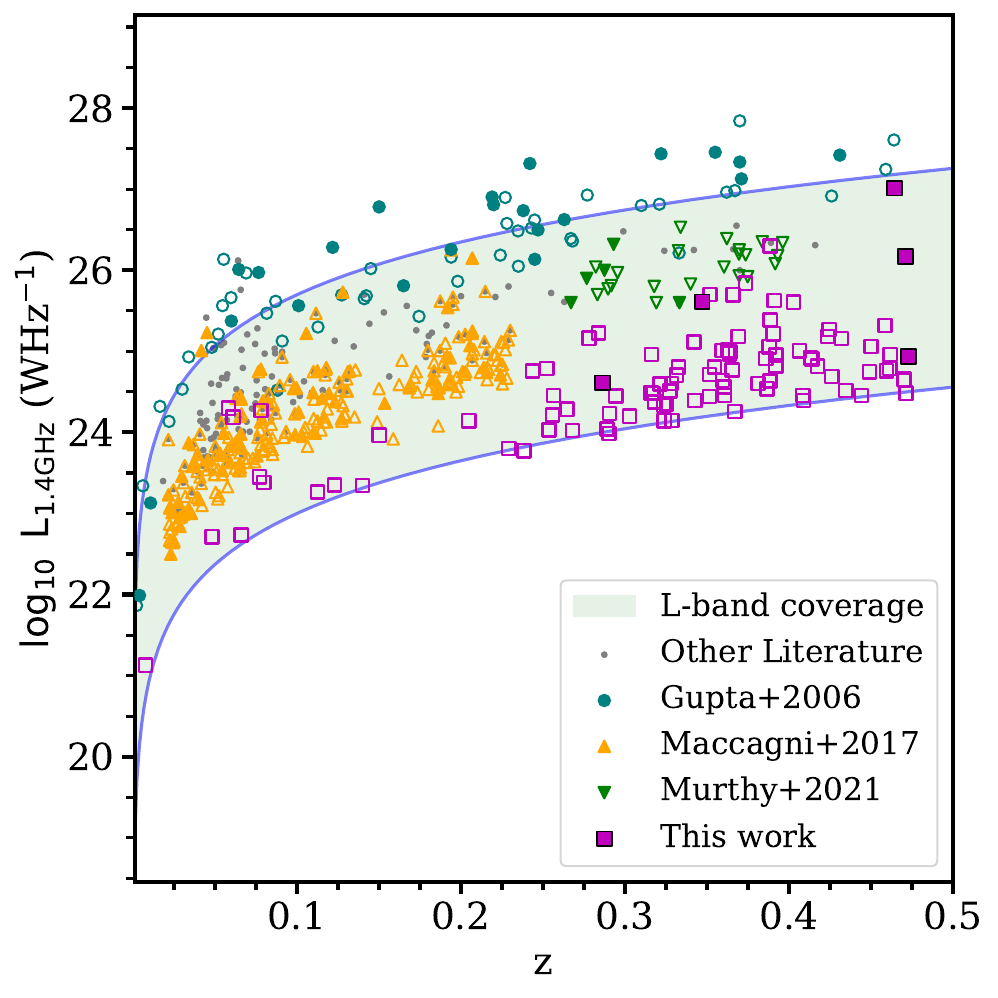}  
}} 
}}  
\vskip+0.0cm   
\caption{L$_{\rm 1.4\,GHz}$ vs $z$ at $z<0.5$ for MALS HERG/LERGs, \cite{Gupta06}, WSRT \citep[][]{Maccagni14} and JVLA \citep[][]{Murthy2021} samples. Filled and open symbols represent \hi\ detections and non-detections, respectively. Other literature sources shown for context include \cite{Glowacki2017, Chandola2017, Allison12, Aditya18gps, Aditya18, Chandola2020, Murthy2022}. The shaded region denotes the MALS L-band search space, where the lower boundary corresponds to the 4\,mJy flux density limit and the upper boundary represents a 2~Jy arbitrary flux density limit.  
}
\label{fig:pow}
\end{figure}

The {\it bottom row} of Fig.~\ref{fig:lzmir} shows measurements from the WSRT survey of \citet{Maccagni17}.  These measurements have been scaled uniformly for a line width (velocity resolution) of 20\,\kms.  
Before comparison, we note that the WSRT survey was pointed observations of preselected bright sources ($>$30\,mJy) at $0.02<z<0.25$, typically reaching 3$\sigma$ optical depth sensitivity of 1\% per channel width ($16$\,\kms). They reported an overall detection rate of $27\pm6$\%. MALS, as an untargeted survey, provides a representative sample down to much fainter flux densities (4\,mJy) but at a lower overall sensitivity. To enable comparison between these two surveys with different sensitivities, detection rates for a range of optical depth sensitivities are provided in the bottom half of Table~\ref{tab:detrates} (see also Fig.~\ref{fig:pow}).
Our $3-5$\% detection rate for high-$\tau$ population ($\int\tau$dv $>$ 5\,\kms) is consistent with the detection rate of \cite{Maccagni17}.  This implies no  significant redshift evolution in the properties of cold gas as probed by \hi\ 21-cm absorption. Notably, when the two samples are matched in 1.4\,GHz luminosity and $\int\,\tau_{3\sigma}\,dv$, the overall detection rates are $8.3^{+19.2}_{-6.9}$\% (1/12) in MALS sample and $20.8^{+4.5}_{-3.8}$\%(30/144) in \cite{Maccagni17} sample,  which are again consistent within the uncertainties.  
%

\citet[][]{Murthy2021} detected \hi\, absorption in $19\%$ of 26 radio loud AGN at $0.25<z<0.4$ with the JVLA. Their sample was drawn exclusively from the bright, high-power end of the radio luminosity function ($400-800$\,mJy; $>25.7$~W\,Hz$^{-1}$).
To ensure consistency with their reported sensitivities, we smoothed the MALS spectra to a velocity resolution of 50\,\kms.  Although the MALS and VLA samples have similar redshift distributions (KS test, $p$-value = 0.12), they differ significantly in radio luminosity ($p$-value = $4.2 \times 10^{-19}$). The VLA sample, as expected, is significantly more luminous (median $\log \rm L_{\rm 1.4\,GHz}\,(\rm{W\,Hz^{-1}}) = 26.02$) compared to the MALS sample (median $\log \rm L_{\rm 1.4\,GHz}\,(\rm{W\,Hz^{-1}}) = 24.64$).  Nevertheless, the detection rates are consistent within uncertainties across all optical depth thresholds  over the luminosity range $\log \rm L_{1.4\,\rm GHz}$ (W\,Hz$^{-1}$) $\sim 21.1 - 27.0$. For instance, at $\int \tau_{3\sigma}\,dv < 5$\,\kms, the detection rates are $6.4^{+6.2}_{-3.5}\%$ and $15.39^{+12.2}_{-7.4}\%$ for MALS and VLA samples, respectively.  This further reinforces our finding that \hi\ 21-cm detection rates show no significant dependence on either 1.4-GHz radio luminosity or redshift. 
Notably, \citet[][]{Murthy2021} found no evidence of fast \hi\ outflows ($>$1000\,\kms). Our two blueshifted detections show moderate velocity offsets rather than extreme broad-wing outflow signatures.

\section{Summary}
\label{sec:summary_LERGHERG}

We presented results from a search for cold neutral gas associated with radio-loud AGNs at $z < 0.5$, which spans $\sim 36$\% of cosmic history, using \hi\ 21-cm absorption line measurements from the MeerKAT Absorption Line Survey (MALS). We cross-matched the MALS DR1 radio continuum catalog at 1006\,MHz with the SDSS DR18 spectroscopic catalog to identify 1908 radio sources, of which 681 were classified as AGNs using BPT diagnostics and the H$\alpha$ versus 1.4\,GHz radio luminosity criterion. Optical emission-line fluxes were measured using the GELATOv2.5.2 software,  and multiple line-ratio diagnostics combined with rest equivalent widths of [O~{\sc iii}]$\lambda$5007 were used to classify 426 of these AGNs into 327 LERGs and 99 HERGs. LERGs showed significantly redder ($>$3$\sigma$) $k$-corrected $g - r$ optical colors, consistent with being hosted by older, more massive galaxies with suppressed star formation. HERGs exhibited slightly steeper median radio spectral indices, reflecting a higher prevalence of FRII-type, lobe-dominated morphologies.  When matched with the \citet[][]{Best12} catalog in 1.4\,GHz luminosity bins, no statistically significant redshift dependence in the overall LERG fraction was found.

We searched for \hi\, 21-cm absorption in a radio-bright subsample of 99 AGNs (79 LERGs, 20 HERGs) with peak flux densities $S_{1.4\text{GHz}} > 4$~mJy,  and reported five new detections (four LERGs and one HERG). 
The sample spanning nearly six decades in radio luminosity (log\,L$_{\rm 1.4GHz}$\,(W\,Hz$^{-1}$) $\sim$ 21.1-27.0), reaching luminosities an order of magnitude fainter than previous targeted \hi\ absorption surveys.
After accounting for varying optical depth sensitivities and the survey completeness, the overall detection rate was $3_{-2}^{+3}$\% at a $3\sigma$ integrated optical depth sensitivity threshold of $10.0$~\kms. 
Comparison with the lower-redshift ($z < 0.25$) survey by \cite{Maccagni17}, matched in radio luminosity and integrated optical depth sensitivity, yielded consistent detection rates, suggesting little to no redshift evolution in the incidence of cold atomic gas out to $z \sim 0.5$.  
Comparison with the VLA survey of \cite{Murthy2021}, which targeted significantly more luminous radio sources ($\rm L_{1.4\text{GHz}} \sim 10^{25.7}-10^{26.5}$~W\,Hz$^{-1}$) over a similar redshift range,  yielded consistent detection rates across all optical depth thresholds, implying no significant dependence of the \hi\ 21-cm detection rate on 1.4\,GHz radio luminosity across $\log \rm L_{1.4,\mathrm{GHz}}$~(W\,Hz$^{-1}$) $\sim 21.1-27.0$.

The five absorbers exhibited complex and disturbed gas kinematics. In three systems the absorption was entirely redshifted relative to the systemic velocity, indicative of inflowing gas.  
In J0943-0004, the merger-driven origin of infalling gas is supported by the evidence of a nearby companion galaxy and double-peaked [O~{\sc iii}]$\lambda\lambda$4959,5007 emission line. In J1455+0441, the absorption occurs asymmetrically against a lobe hotspot, potentially tracing gas from a faint companion galaxy  at a projected distance of $\sim 150$~kpc from the core.
In two systems, J0228-0107 and J1117-0235, the absorption was fully or predominantly blueshifted, with  high-resolution radio morphology suggesting jet-driven outflows.
\hi\, profiles associated with LERGs displayed a wide range of asymmetries ($A_p$ = 0.07–0.65) and velocity offsets exceeding 350\,\kms, indicative of cold gas disturbed by jet–ISM interactions or lobe expansion, consistent with findings from lower-$z$ samples \citep[e.g.][]{Glowacki2017, Chandola2020}.  
The single HERG detection precluded a rigorous kinematic comparison between the two accretion modes.
Future work will extend this study to the full MALS AGN sample and to star-forming galaxies by combining MALS with upcoming spectroscopic releases from DESI \citep[][]{DESIDR12025} and 4MOST surveys \citep[e.g.,][]{deJong2019, Krogager23_PAQS}.


\begin{acknowledgements}
The MeerKAT telescope is operated by the South African Radio Astronomy Observatory (SARAO; \url{www.sarao.ac.za}), which is a facility of the National Research Foundation, an agency of the Department of Science and Innovation. PPD hereby acknowledges the financial assistance of SARAO towards this research. The MeerKAT data were processed using the MALS computing facility at IUCAA (\url{https://mals.iucaa.in/releases}).
The National Radio Astronomy Observatory is a facility of the National Science Foundation operated under cooperative agreement by Associated Universities, Inc.  
This research has made use of NASA's Astrophysics Data System and the NASA/IPAC Extragalactic Database (NED), which is operated by the Jet Propulsion Laboratory, California Institute of Technology, under contract with the National Aeronautics and Space Administration. 
\end{acknowledgements}

\def\aj{AJ}%
\def\actaa{Acta Astron.}%
\def\araa{ARA\&A}%
\def\apj{ApJ}%
\def\apjl{ApJ}%
\def\apjs{ApJS}%
\def\ao{Appl.~Opt.}%
\def\apss{Ap\&SS}%
\def\aap{A\&A}%
\def\aapr{A\&A~Rev.}%
\def\aaps{A\&AS}%
\def\azh{AZh}%
\def\baas{BAAS}%
\def\bac{Bull. astr. Inst. Czechosl.}%
\def\caa{Chinese Astron. Astrophys.}%
\def\cjaa{Chinese J. Astron. Astrophys.}%
\def\icarus{Icarus}%
\def\jcap{J. Cosmology Astropart. Phys.}%
\def\jrasc{JRASC}%
\def\mnras{MNRAS}%
\def\memras{MmRAS}%
\def\na{New A}%
\def\nar{New A Rev.}%
\def\pasa{PASA}%
\def\pra{Phys.~Rev.~A}%
\def\prb{Phys.~Rev.~B}%
\def\prc{Phys.~Rev.~C}%
\def\prd{Phys.~Rev.~D}%
\def\pre{Phys.~Rev.~E}%
\def\prl{Phys.~Rev.~Lett.}%
\def\pasp{PASP}%
\def\pasj{PASJ}%
\def\qjras{QJRAS}%
\def\rmxaa{Rev. Mexicana Astron. Astrofis.}%
\def\skytel{S\&T}%
\def\solphys{Sol.~Phys.}%
\def\sovast{Soviet~Ast.}%
\def\ssr{Space~Sci.~Rev.}%
\def\zap{ZAp}%
\def\nat{Nature}%
\def\iaucirc{IAU~Circ.}%
\def\aplett{Astrophys.~Lett.}%
\def\apspr{Astrophys.~Space~Phys.~Res.}%
\def\bain{Bull.~Astron.~Inst.~Netherlands}%
\def\fcp{Fund.~Cosmic~Phys.}%
\def\gca{Geochim.~Cosmochim.~Acta}%
\def\grl{Geophys.~Res.~Lett.}%
\def\jcp{J.~Chem.~Phys.}%
\def\jgr{J.~Geophys.~Res.}%
\def\jqsrt{J.~Quant.~Spec.~Radiat.~Transf.}%
\def\memsai{Mem.~Soc.~Astron.~Italiana}%
\def\nphysa{Nucl.~Phys.~A}%
\def\physrep{Phys.~Rep.}%
\def\physscr{Phys.~Scr}%
\def\planss{Planet.~Space~Sci.}%
\def\procspie{Proc.~SPIE}%
\let\astap=\aap
\let\apjlett=\apjl
\let\apjsupp=\apjs
\let\applopt=\ao

\bibliographystyle{aa}
\bibliography{mybib}

\begin{appendix}

\section{Representative optical spectra and Diagnostic framework}
\label{sec:optical_fit}

 In this section, we present representative fits to the optical spectra performed with {\tt GELATO} and describe the diagnostic framework used to classify AGNs.

\begin{figure*}[h]
\centerline{\vbox{
\centerline{\hbox{ 
\includegraphics[
trim = {0cm 0cm 0cm 0cm}, clip=true,
width=0.90\textwidth,angle=0]{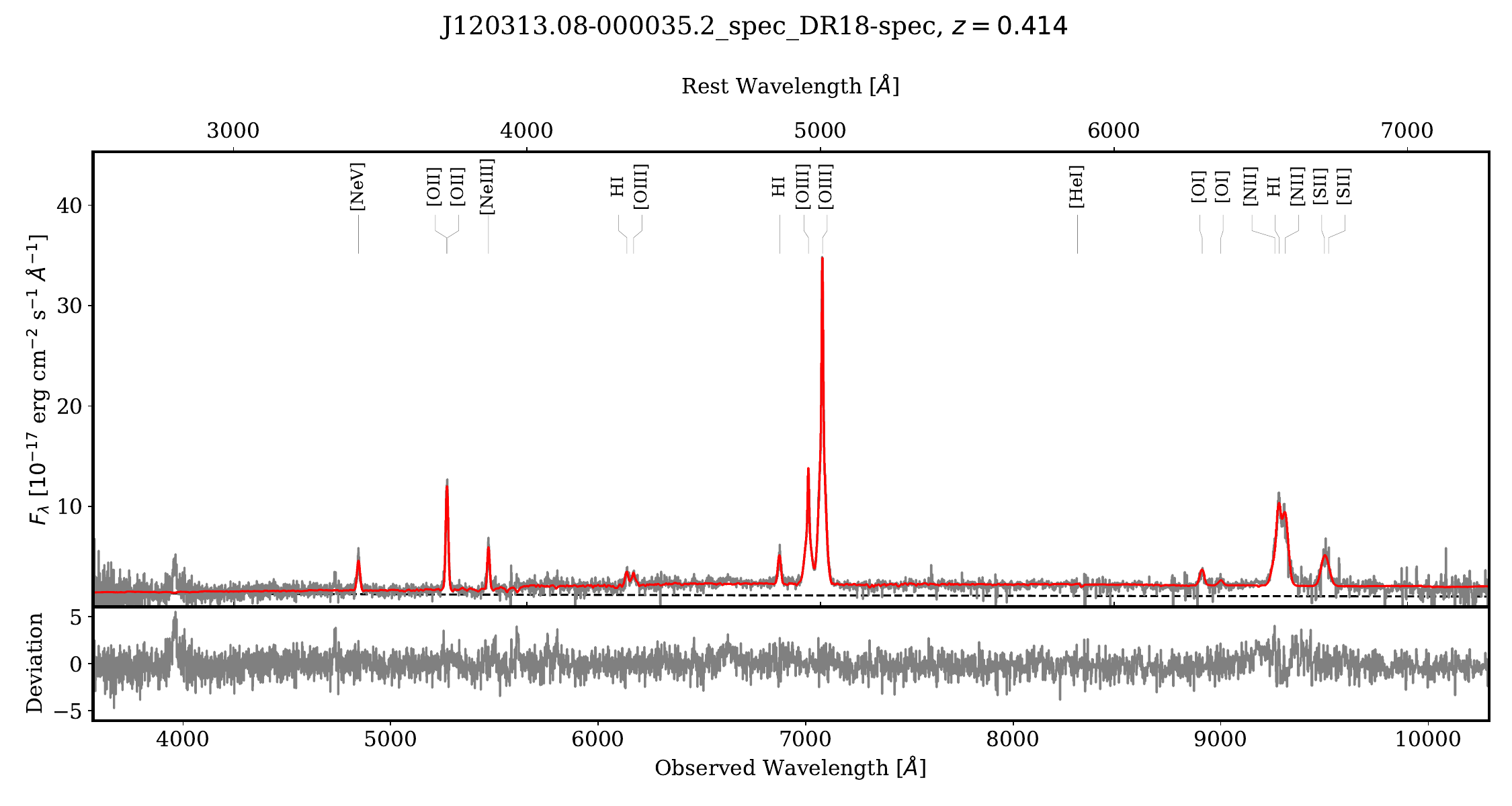}  
}} 
}}
\centerline{\vbox{
\centerline{\hbox{ 
\includegraphics[
trim = {0cm 0cm 0cm 0cm}, clip=true,
width=0.90\textwidth,angle=0]{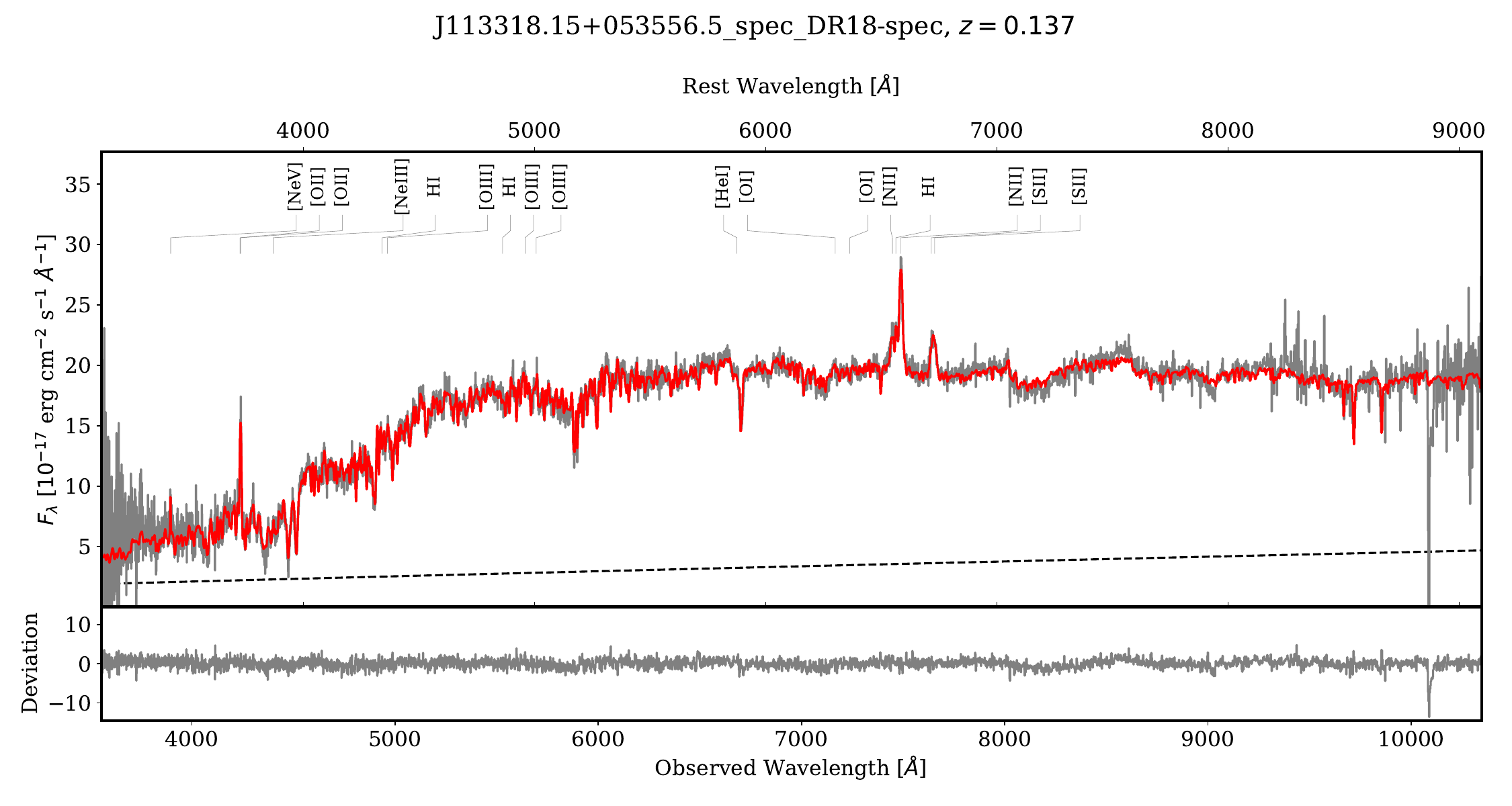}  
}} 
}}
\vskip+0.0cm   
\caption{Representative spectra classified via the Excitation Index ({\tt Diagnostic 1}), showing a  HERG ({\it Top}) and LERG ({\it bottom}). In each panel, the upper sub-panel shows the SDSS spectrum (gray), the {\tt GELATO} fit (red), and the positions of emission-line components (Table~\ref{tab:GELATO_PARAMS}). The lower sub-panel shows the residuals, normalized by error.
}
\label{fig:EI}
\end{figure*}

\begin{figure*}[h]
\centerline{\vbox{
\centerline{\hbox{ 
\includegraphics[
trim = {0cm 0cm 0cm 0cm}, clip=true,
width=0.90\textwidth,angle=0]{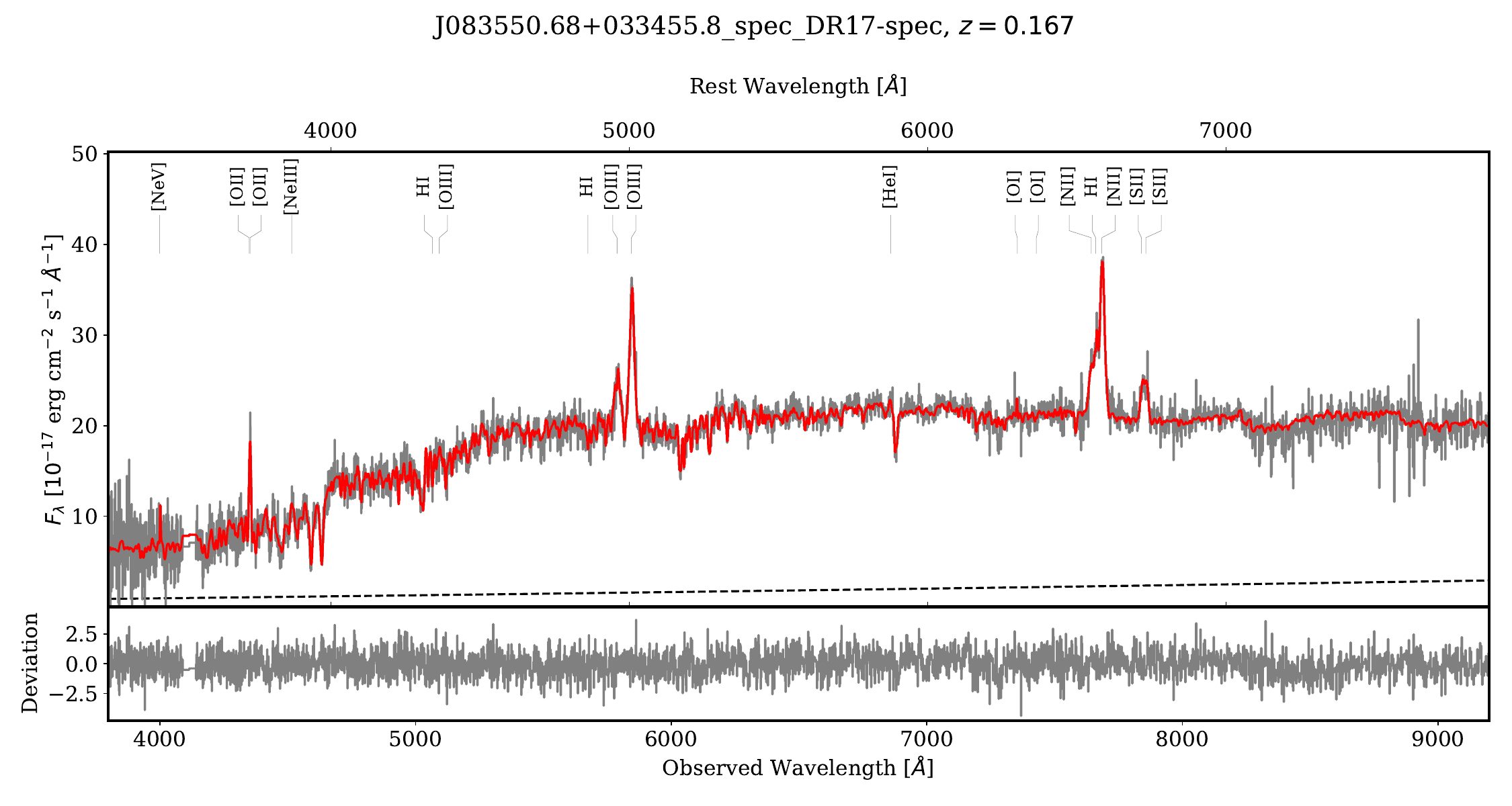}  
}} 
}}
\centerline{\vbox{
\centerline{\hbox{ 
\includegraphics[
trim = {0cm 0cm 0cm 0cm}, clip=true,
width=0.90\textwidth,angle=0]{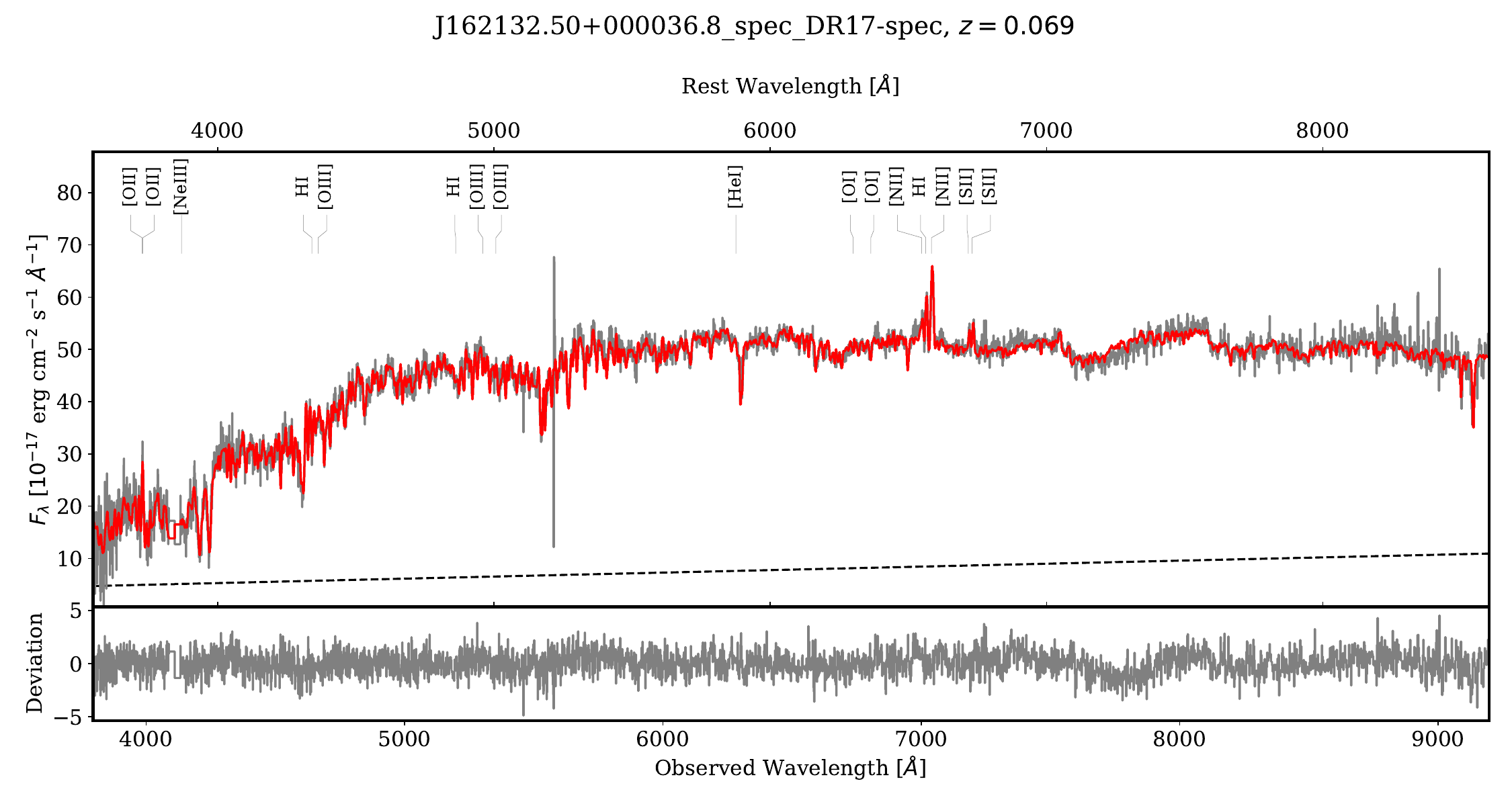}  
}} 
}}
\vskip+0.0cm   
\caption{Representative spectra classified using the diagnostic diagrams from \cite{Kewley06} ({\tt Diagnostic 2}).
}
\label{fig:K06}
\end{figure*}

\begin{figure*}[h]
\centerline{\vbox{
\centerline{\hbox{ 
\includegraphics[
trim = {0cm 0cm 0cm 0cm}, clip=true,
width=0.90\textwidth,angle=0]{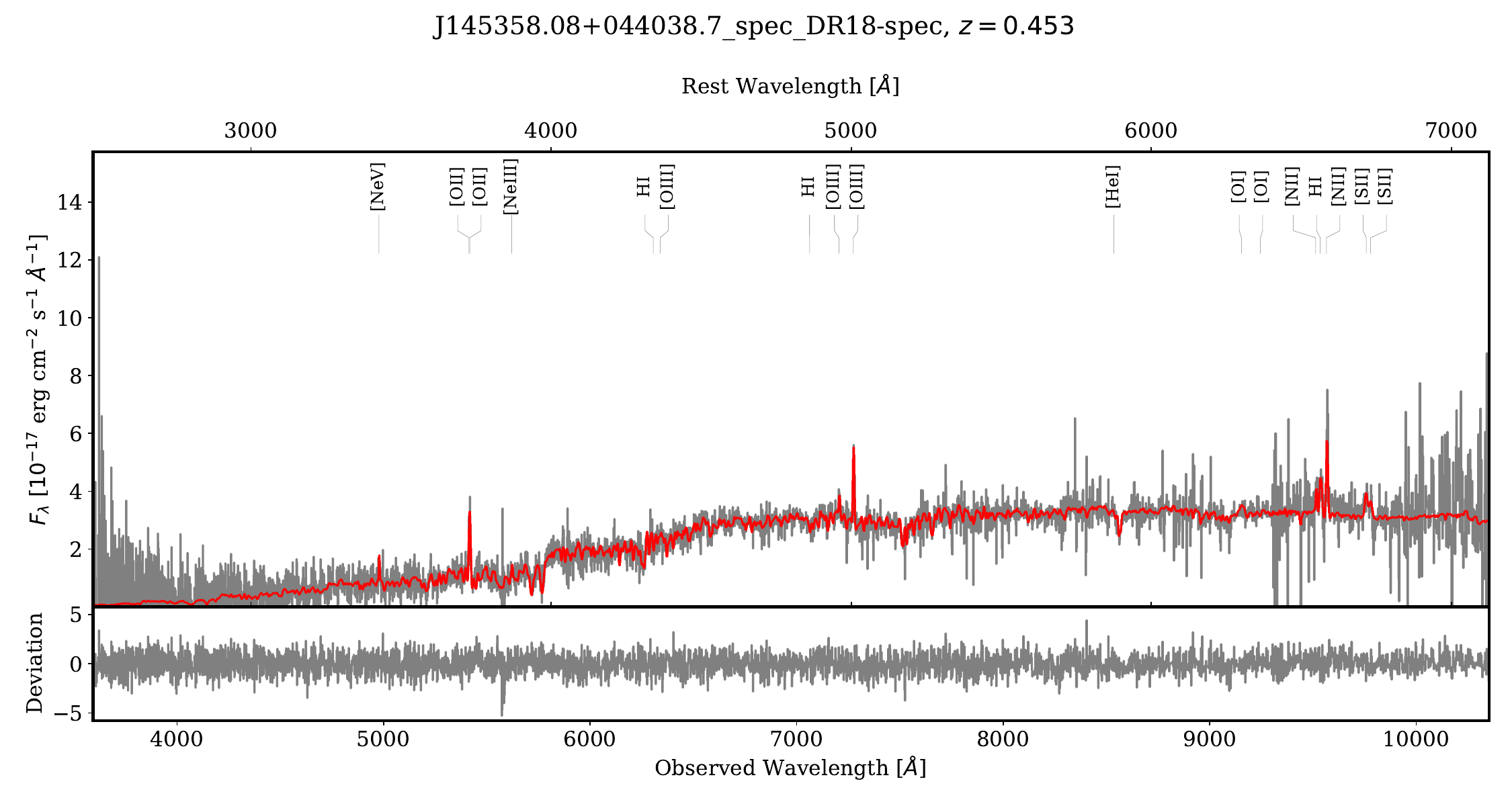}  
}} 
}}
\centerline{\vbox{
\centerline{\hbox{ 
\includegraphics[
trim = {0cm 0cm 0cm 0cm}, clip=true,
width=0.90\textwidth,angle=0]{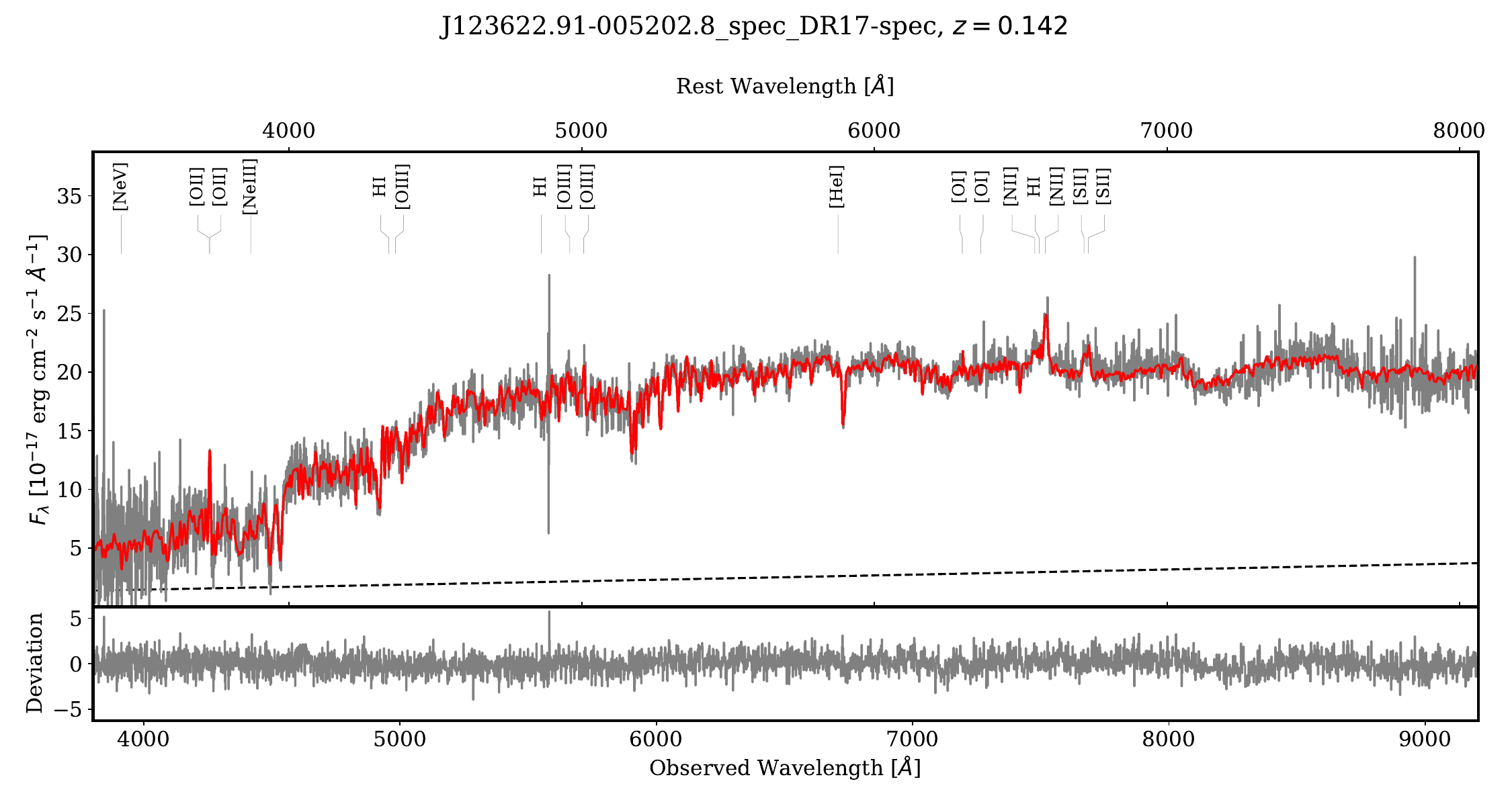}  
}} 
}}
\vskip+0.0cm   
\caption{Representative spectra classified using the \oiii\,$\lambda 5007$ rest-frame equivalent width  ({\tt Diagnostic 3}).
}
\label{fig:OIIIEW}
\end{figure*}

\begin{figure*}[h]
\centerline{\vbox{
\centerline{\hbox{ 
\includegraphics[
trim = {0cm 0cm 0cm 0cm}, clip=true,
width=0.90\textwidth,angle=0]{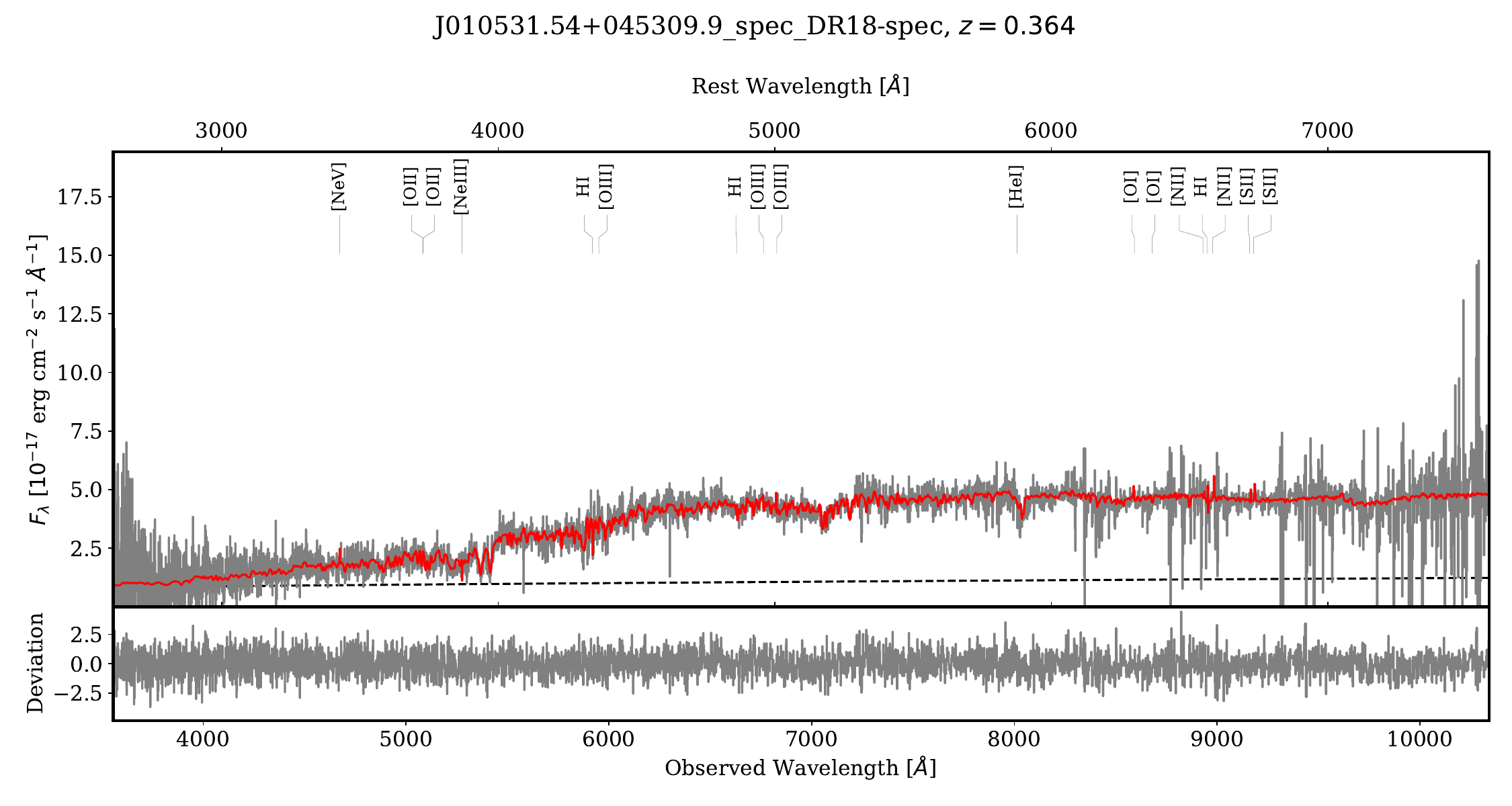}  
}} 
}}
\centerline{\vbox{
\centerline{\hbox{ 
\includegraphics[
trim = {0cm 0cm 0cm 0cm}, clip=true,
width=0.90\textwidth,angle=0]{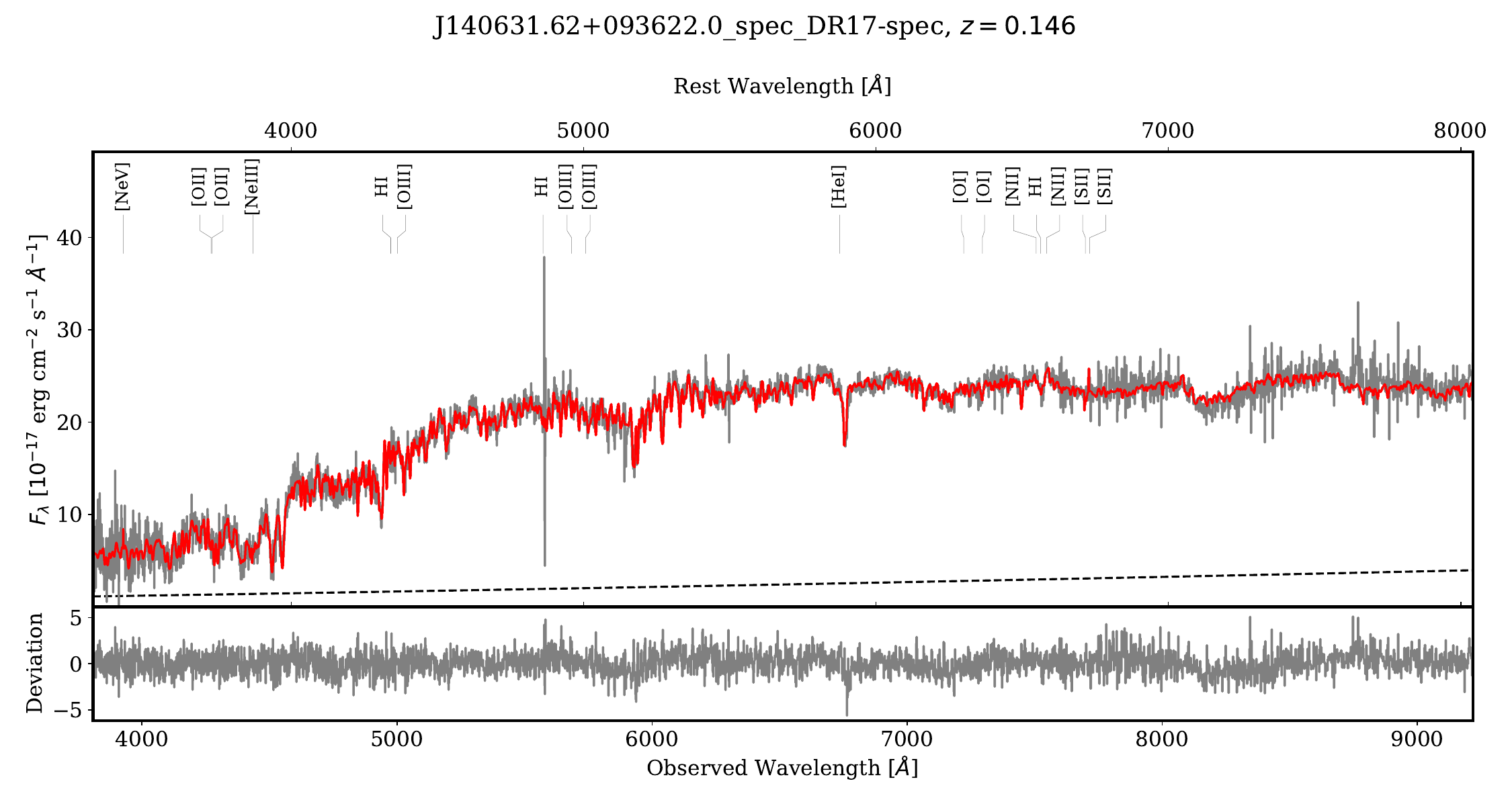}  
}} 
}}
\vskip+0.0cm   
\caption{Representative spectra classified using the diagnostic diagrams from \cite{CidFernandes10} ({\tt Diagnostic 5}).
}
\label{fig:CF10}
\end{figure*}

An AGN may be classified as LERG or HERG based on the set of five diagnostics listed in Section~\ref{sec:LERG_HERG}.  For each diagnostic, we define a confidence measure $C$ depending on following three factors, $f_i$, with $|f_i| \in [0, 1]$):
\begin{enumerate}
    \item \textbf{Line Quantity Factor} ($f_1$): This reflects the number of emission lines used in the classification,
    \[
    f_1 =  \frac{N_{\mathrm{lines}} - N_{\mathrm{min}}}{N_{\mathrm{max}} - N_{\mathrm{min}}},
    \]
    where $N_{\mathrm{max}} = 6$ and $N_{\mathrm{min}} = 1$.
    
    \item \textbf{S/N Factor} ($f_2$): This accounts for the minimum S/N used among the lines,
    \[
    f_2 = \frac{1}{1 + \exp[-k_1(\mathrm{S/N}_{\mathrm{min}} - \mathrm{S/N}_0)]},
    \]
    with $\mathrm{S/N}_0 = 3$ as a reference threshold and we assume $k_1 = 2$, that controls the steepness of the function.

    \item \textbf{Distance from Boundary} ($f_3$): This measures distance of the source in the units of uncertainty from the  diagnostic-specific segregation boundary,
    \[
    f_3 = \frac{1}{1 + \exp[-k_2(\Delta - 1)]},
    \]
    where $\Delta = |x - x_{\mathrm{div}}|/\sigma$ and steepness parameter $k_2$ = 2.
\end{enumerate}
The confidence measure, C, for a diagnostic is the weighted average of $f_i$'s with weights $w_i$ as C = $\sum$ $w_i$$f_i$, where $i$ = 1, 2 and 3.  We assume equal weights, $w_1 = w_2 = w_3 = \frac{1}{3}$, to assign equal importance to all three factors. We assign
a negative sign for LERGs and a positive sign for HERGs.
Each diagnostic is also assigned a reliability factor $R \in [0, 1]$. 
The final significance ($S$) for the classification of a source is then calculated as,
\begin{equation}\label{eq:S}
    S = \frac{\sum_{i=1}^5 R_i\cdot C_i}{\sum_{i=1}^5 R_i},
\end{equation}
%

The EI-based diagnostic ({\tt Diagnostic 1}) is inapplicable to AGNs with broad Balmer components \citep[][]{Buttiglione10}. Therefore, to identify optimal weights, we used 433 narrow-line AGNs (Table~\ref{tab:GELATO_PARAMS}). 
A sequential application of these diagnostics identifies: 11 HERGs/ 21 LERGs ({\tt Diagnostic 1}), 10 HERGs / 22 LERGs ({\tt Diagnostic 2}),  $27$ HERGs / $107$ LERGs ({\tt Diagnostic 3}),  10 HERGs / 8 LERGs ({\tt Diagnostic 4}) and  $13$ HERGs / $147$ LERGs ({\tt Diagnostic 5}). Clearly, more objects are classified as subsequent less stringent diagnostics are used.  
 In total, 376 out of 531 AGNs get classified using at least one diagnostic method in  to $70$ HERGs and $306$ LERGs.

\begin{figure}[t]
\centerline{\vbox{
\centerline{\hbox{ 
\includegraphics[
trim = {0cm 0cm 0cm 0cm}, clip=true,
width=0.50\textwidth,angle=0]{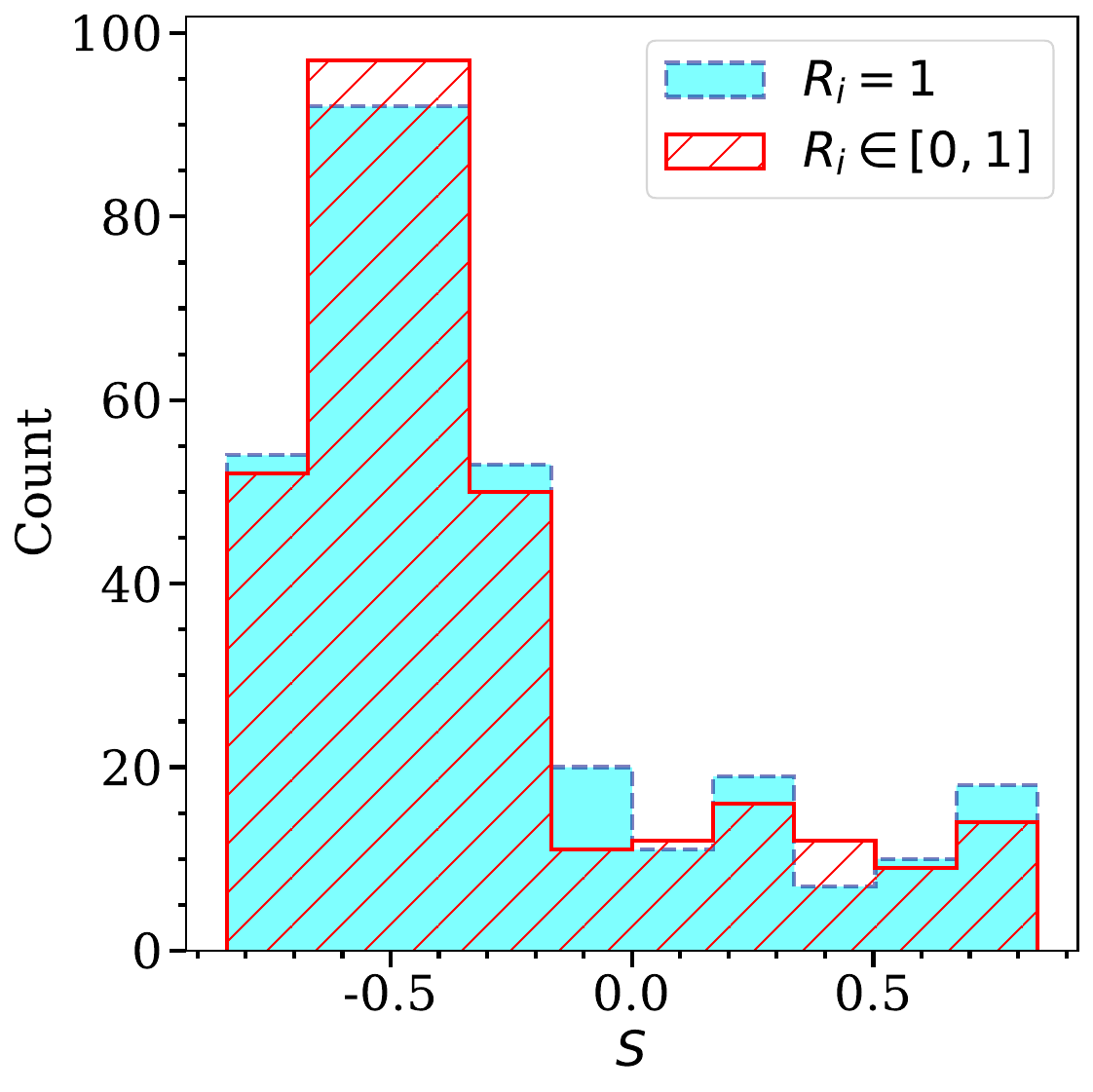}  
}} 
}}  
\vskip+0.0cm   
\caption{Distribution of significance scores ($S$) for LERG/HERG classification. The cyan-filled histogram represents equal weights ($R_i = 1$), while the red-hatched histogram shows the distribution using the diagnostic-specific weights -- $R_1 = 1$, $R_2 = R_3 = 0.9$, $R_4 = 0.8$, and $R_5 = 0.5$.}
\label{fig:diag-scores}
\end{figure}

In comparison to the above, adopting uniform weights ($R_i$ = 1), implying no preference for any diagnostic, we get 60 HERGs and 236 LERGs. However, to better align with sequential classification used in the literature, we adopted $R_1=1$, $R_2= R_3 = 0.9$, $R_4=0.8$ and $R_5=0.5$.  The distribution of significances are shown in Fig.~\ref{fig:diag-scores}.  
This yields 68 HERGs and 308 LERGs, with a few notable exceptions highlighting the objectivity of our approach.  
For example, J091103.75$+$191752.0 $-$ was initially classified as a HERG based solely on \oiii\, REW of $9.5 \pm 0.8$~\AA, but our final significance score ($S \sim -0.21$) correctly identifies it as a LERG (see also last paragraph of Section~\ref{sec:LERG_HERG}).

It is  particularly intriguing to test the classifications  of 160 AGNs initially categorized using only weak-line diagnostics of \citet[][]{CidFernandes10} ({\tt Diagnostic 5}). 
To facilitate this, we lowered the line flux S/N threshold to 1 to incorporate additional diagnostics where possible. For 10 sources -- J101133.55+155812.9 ($S\sim -0.20$), J012858.28+141337.7 ($S\sim -0.25$), J001450.80+082545.0 ($S\sim -0.20$), J002205.85+060317.1 ($S\sim -0.14$), J010531.54+045309.9 ($S\sim -0.23$), J130746.13+080108.1 ($S\sim -0.17$), J083821.14+034405.7 ($S \sim -0.19$), J230225.22+055418.2 ($S \sim -0.39$), J092728.26+020438.6 ($S \sim -0.37$), and J225804.30$-$102143.7 ($S \sim -0.43$) -- the resulting significance scores indicated a different classification than before. 

\section{MALS-SDSS sample}

Table~\ref{tab:lerg_herg_samp} presents the first 10 rows of the MALS-SDSS sample of 1300 sources. Of these, 426 are classified into LERGs (327) and HERGs (99), with the full list available in the electronic version.

\begin{table*}[]
\setlength{\tabcolsep}{6pt}
\caption{MALS-SDSS sample classified as SF or composite galaxies and AGN (LERGs/HERGs). \label{tab:lerg_herg_samp}} 
\begin{tabular}{lcccccccc}
\hline\hline
Source  & MALS RA  & MALS DEC & $z$ & S$_{1.4\rm GHz}$ & $\log\rm L_{1.4\rm GHz}$ &  Class & $S$ & Spectroscopic Class \\
        & (deg)    & (deg)    &     & (mJy)            & (W\,Hz$^{-1}$)           &        &     &                     \\
(1) & (2) & (3) & (4) & (5) & (6) & (7) & (8) & (9)\\ 
\hline
J000002.61$+$200307.3 & 0.01091 & 20.05205 & 0.279 & 0.7 & 23.33 & COMP & \nodata & \nodata \\
J000334.74$-$050307.4 & 0.89476 & $-5.05208$ & 0.499 & 1.8 & 24.22 & AGN &  \nodata & \nodata \\
J000337.42$-$053401.6 & 0.90595 & $-5.56711$ & 0.396 & 3.1 & 24.21 & AGN &  \nodata & \nodata \\
J000432.52$-$045650.9 & 1.13554 & $-4.94748$ & 0.243 & 13.7 & 24.37 & AGN & \nodata & \nodata \\
J000449.86$+$174303.4 & 1.20777 & 17.71763 & 0.446 & 1.7 & 24.07 & AGN & $-0.17$ & LERG \\
J000519.61$-$050810.3 & 1.33174 & $-5.13622$ & 0.241 & 3.1 & 23.71 & AGN &  \nodata & \nodata \\
J000747.87$+$165900.9 & 1.94949 & 16.98359 & 0.403 & 6.3 & 24.57 & AGN & \nodata & \nodata \\
J000756.44$+$175635.5 & 1.98517 & 17.94320 & 0.453 & 3.8 & 24.44 & AGN & 0.84 & HERG \\
J001218.32$+$084415.2 & 3.07634 & 8.73757 & 0.261 & 1.5 & 23.29 & AGN & \nodata & \nodata \\
J001237.48$+$084353.7 & 3.15620 & 8.73159 & 0.165 & 22.4 & 24.23 & AGN & \nodata & \nodata \\
\hline
\hline
\end{tabular}
\textcolor{black}{
\small{
    Column\,1: MALS-SPW2 source ID. Column\,2 -- 3: MALS right ascension and declination, both in degrees. Column\,4: SDSS spectroscopic redshift. Column\,5: flux density at 1.4~GHz in mJy. Column\,6: $\log$ radio luminosity at 1.4~GHz. Column\,7: classification into AGN, SF, and composite (COMP) systems, Column\,8: Significance score as defined in equation~\ref{eq:S}. Column\,9: Optical spectroscopic classification.}  \\
}
\end{table*}

\section{Optical images of AGN with 21-cm absorption detections}

\begin{figure*}[ht]
    \centering
    \begin{minipage}[b]{0.19\textwidth}
        \includegraphics[width=\linewidth]{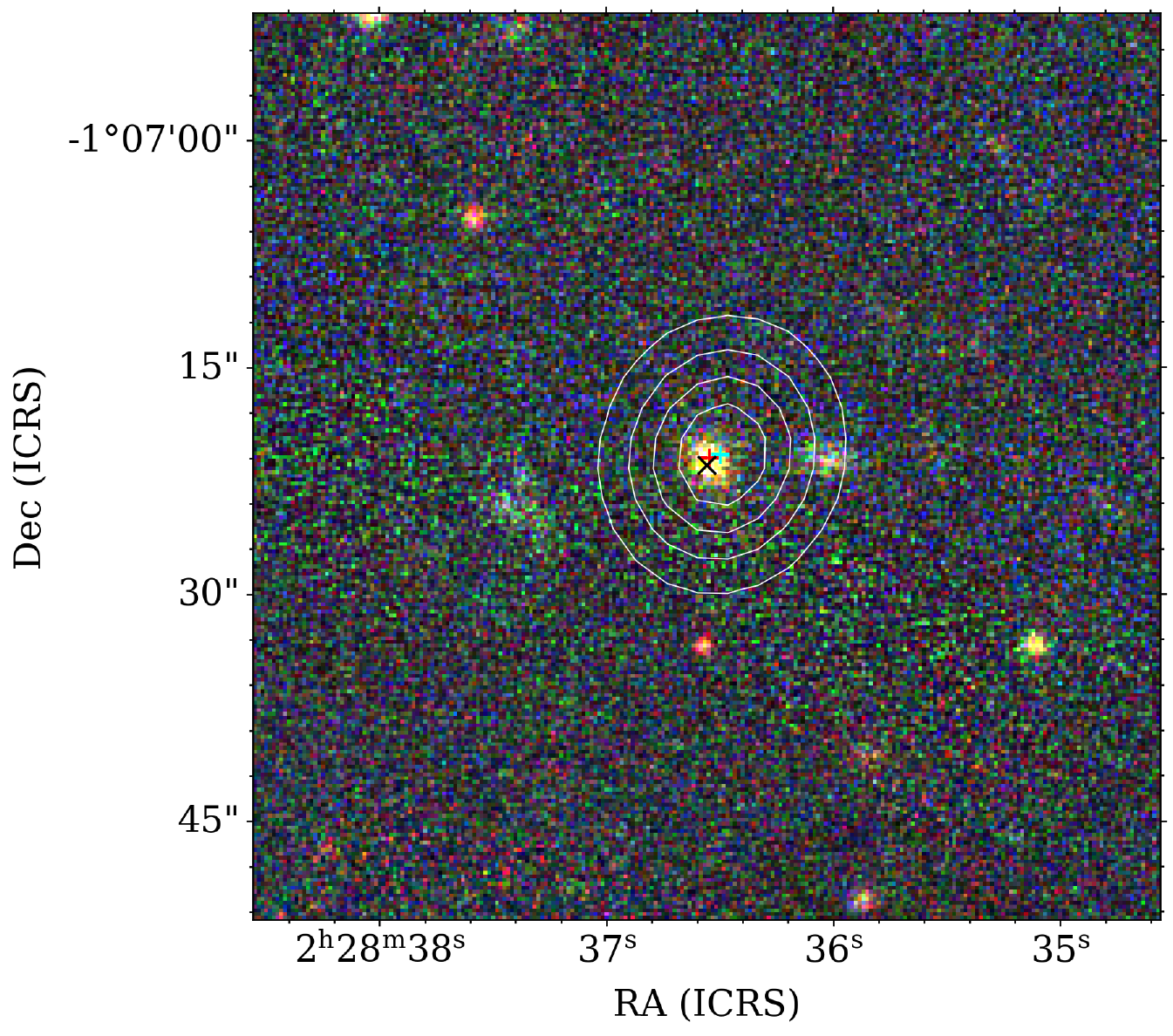}
        \end{minipage}
     \begin{minipage}[b]{0.19\textwidth}
        \includegraphics[width=\linewidth]{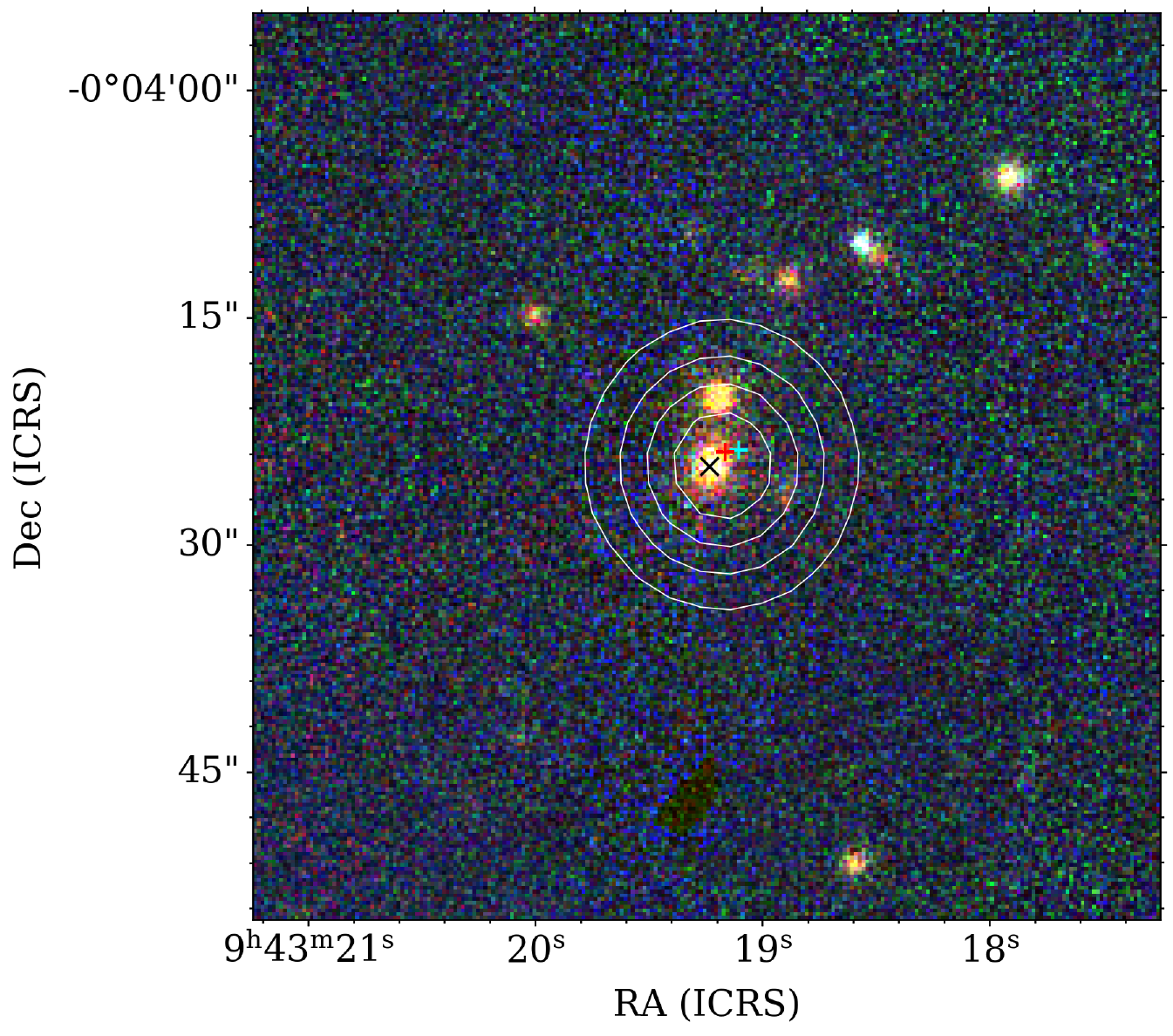}
    \end{minipage}
     \begin{minipage}[b]{0.19\textwidth}
        \includegraphics[width=\linewidth]{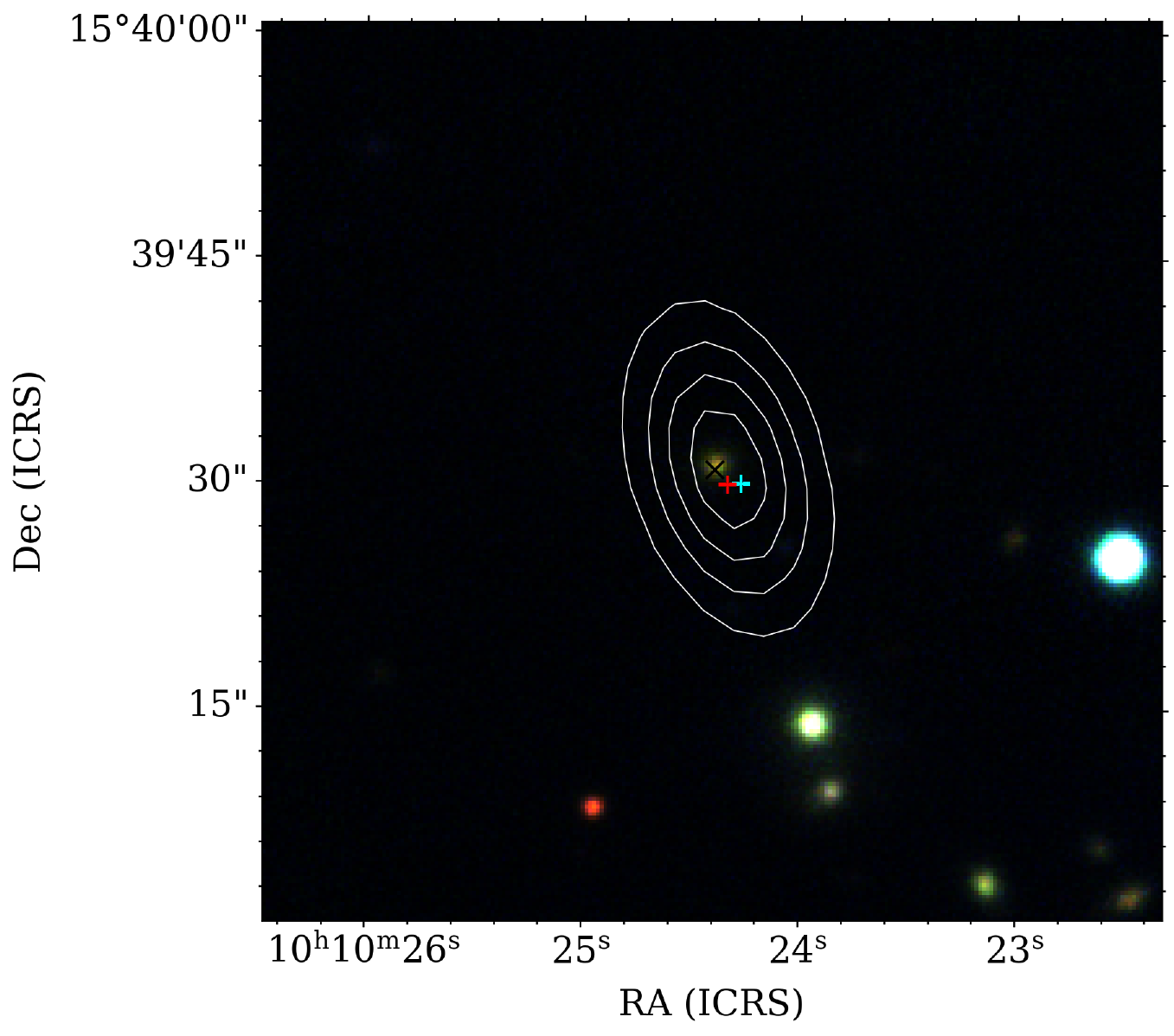}
    \end{minipage}
     \begin{minipage}[b]{0.19\textwidth}
        \includegraphics[width=\linewidth]{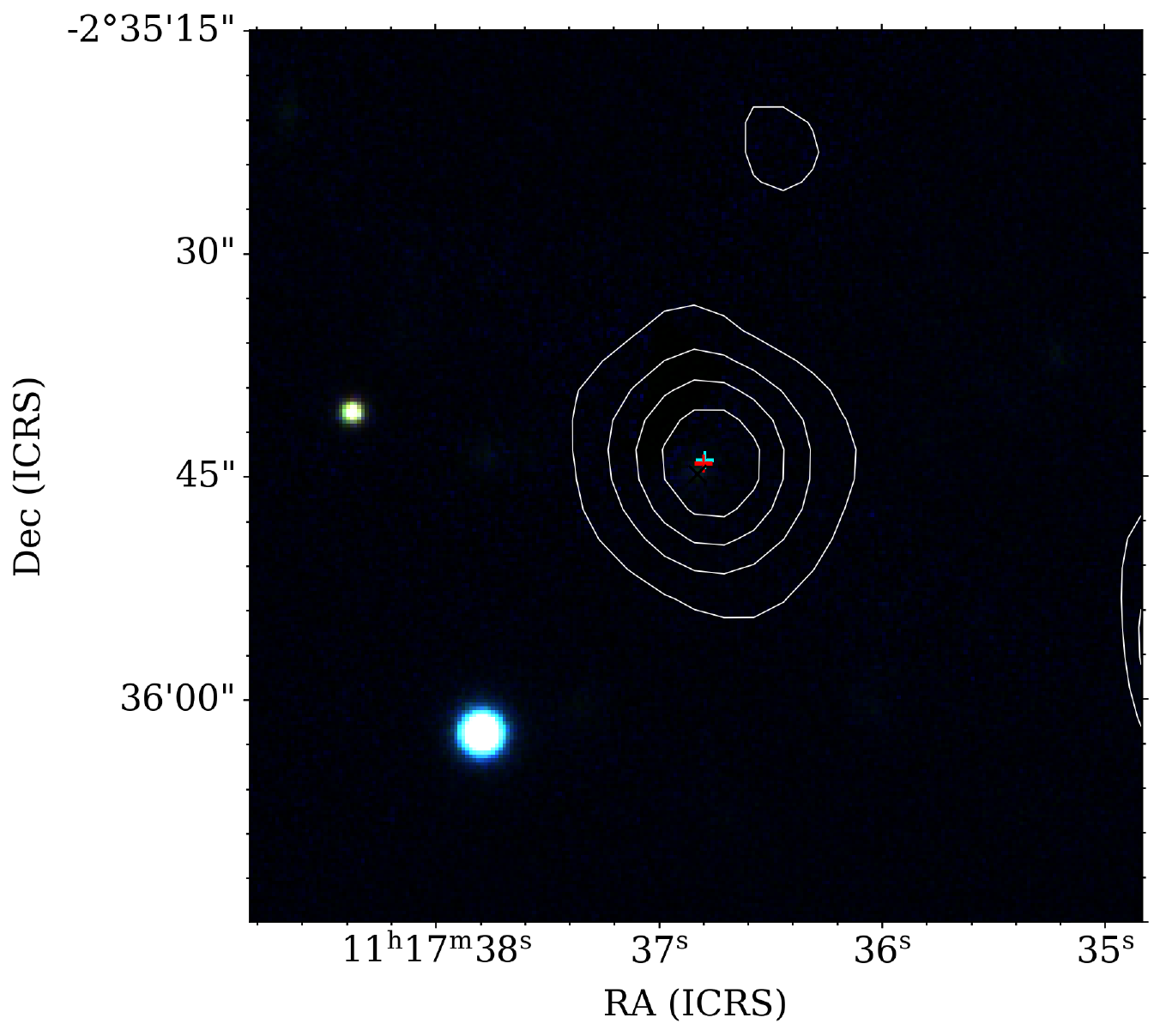}
    \end{minipage}
     \begin{minipage}[b]{0.19\textwidth}
        \includegraphics[width=\linewidth]{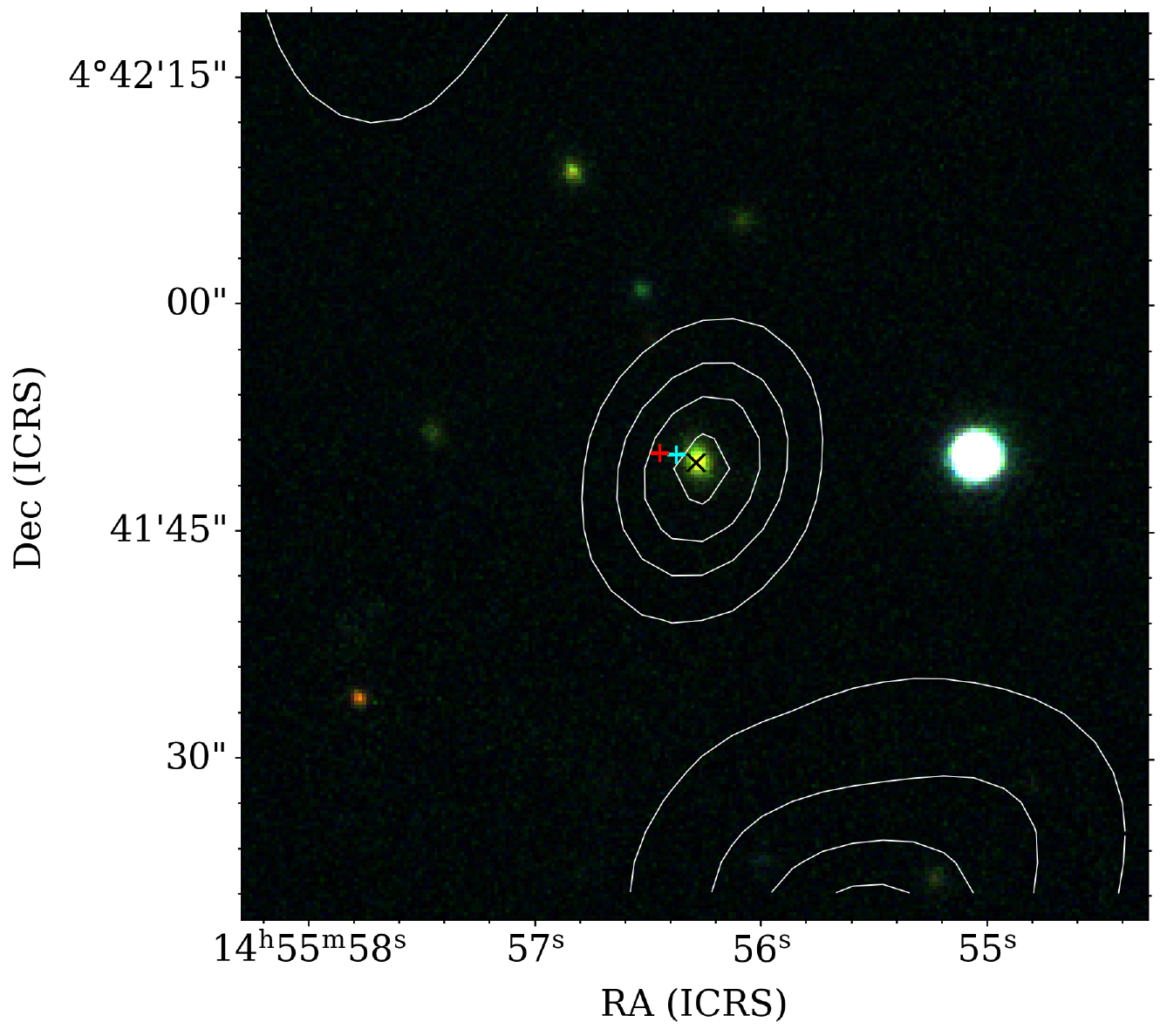}
    \end{minipage}
    \caption{Optical color composite ($g$, $r$, $i$) images from Pan-STARRS1 for the five sources displaying associated \hi\ 21-cm absorption (Table~\ref{tab:detections_2mJy}). From left to right: J022836.49-010720.8, J094319.09-000423.7, J101024.28+153929.9, J111736.77-023544.0, and J145556.40+044150.0.  The contour levels represent 20\%, 40\%, 60\%, and 80\% of the peak flux density of the sources in the radio-band near the \hi\, 21-cm detection frequency.} 
    \label{fig:opt_images}
\end{figure*}

\end{appendix}

\end{document}